\documentclass[a4paper,11pt]{article}
\usepackage{pos}
\usepackage{soul}
\usepackage{amsmath}
\usepackage{adjustbox}
\title{Approximate Minimal SU(5), Several Fundamental Scales, Fluctuating
    Lattice}
  \ShortTitle{Several scales}

\author*[a]{Holger Bech Nielsen}

\affiliation[a]{Niels Bohr Institute,\\
  Jagtvej 155A, Copenhagen N , Denmark}

\emailAdd{hbech@@nbi.dk}

\abstract{Having shortly reviewed our idea of the grand unified SU(5) being
  only exact in a classical limit, in a truly existing lattice, an ontological lattice, we go over to putting a series of different
  physical energy scales such the approximate unification scale for
  the SU(5) (without any SUSY), the Planck scale, and e.g. the scale of
  see-saw neutrino masses into a certain plot showing the energy scales
  on a straght line. This straight line of this plot supposed to result from
  such an
  ontological
  lattice, that fluctuates in link size $a$ and lattice density in a very
  strong way according to a log normal distribution. The point is that
  different energy scales result from the link size of the lattice
  to different powers, like the averages $<a^n>$, where the power $n$ depends 
  on the type of scale considered. Since  the
  $n$'th root of the average of the $n$'th power of the link size in the
  fluctuationg lattice is very strongly dependent on the power $n$, because
  of very huge fluctuations, the different types of physical energy scales
  can get very different.
  
  With a Galton i.e. log normal distribution of the link size the various energy
  scales have their logarithms fall into a nice straight line versus the power
  of the link size, on which they depend.

  Our model gives a surprisingly good
  number
  for the small deviations of the experimental electron- and  muon-anomalous
  magnetic moment
  from the pure
  Standard Model value.}

\FullConference{Proceedings of the Corfu Summer Institute 2024 "School and
  Workshops on Elementary Particle Physics and Gravity" (CORFU2024)\
12 - 26 May, and 25 August - 27 September, 2024\\
Corfu, Greece\\}

\begin{document}
\maketitle

\section{Introduction}

Many physicists would probably speculate, if the Planck scale \cite{PU} should
be
the fundamental energy scale of physics; but there are several other energy
scales of completely different orders of magnitude, the scale of the see-saw
masses e.g. is closer about $10^{11}GeV$, but although unknown by several orders
of magnitude at least not at all equal to the Planck scale. Similarly even in
supersymmetric theories the Grand Unification, say SU(5), scale only hardly
reach the Planck scale, and without SUSY, just taken as the scale of closest
meeting of the three running fine structure constants, the approximate
unification energy scale becomes much smaller than the Planck scale.
So a lot of regimes, with for us to guess new physics, in between these
energy scales, of which we may have hints, seem to have to be present.
(Different ways of looking for fundamental scales may be seen in
\cite{Funkhouser, Stojkovic} and fluctuating or irregular lattices as we
shall use as
our model may be seen in \cite{Vergeles,Wagner}. Our own motivation
for fluctuating lattice is the expectation of fluctuation in the gauge
of gravity, the reparametrization\cite{FNN, trans, RS}.)

In the present article we want to ``unify'' the several energy scales
by proposing a strongly fluctuating lattice, that could provide effectively
{\bf several} order of magnitudewise different scales
(by averaging different powers of the link size). Then the intermediate
regimes would not be needed to connect different energy scales, in as far
as rather it would  all be described by the same cut off
- let us consider it a lattice- which though in our model is fluctuating
and have e.g. different linksizes in different places.

This proceeding is of a talk, that can be considered a review of the recent
article \cite{Universe}, the point of which is that a series of various
energy scales such as the Plack scale\cite{PU}, the grand unified SU(5)
unification scale\cite{AppSU5, GG2}
taken as an only  approximately unification  scale, so that it is only  the
scale, where the running gauge
theory couplings
for the Standard Model- the fine structure constants - 
are closest to agree with the SU(5) relation, etc. can be fitted with
only two parameters. In fact a series of 9 energy scales fall on a straight
line,
if you plot the logarithms of their energy-values versus the integer
number giving, what power of the lattice link length of a truly existing
(ontological fluctuating) lattice, that is related to the  energy scale in
question.

This article \cite{Universe} came out as a follower of my finding in
\cite{AppSU5}
(this work was based on a lot of old works of ours assuming that the gauge
coupling constants were udjusted to be critical couplings for a lattice
(at the Planck scale)\cite{Don137, Don93, RDrel1,RDrel2, RDrel3,Bled8,
  Polonica,LR,
  flipped,LRNwf,
  LRN,crit,Picek}, and the idea of an approximate SU(5) is found in Senjanovic
and Zathedeski
\cite{Senjanovic}.)
of that the deviation of the experimental fine structure constants from the
GUT SU(5) predition could be explained as due to a quantum correction of a
lattice theory, which in the classical approximation is SU(5) invariant
- in the sense of giving the SU(5) relation between the fine structure
constants -,
though with the little trouble: The quantum correction should be just 3 times
bigger than, what the calculation with a simple lattice would give.
Another point of deviation from usual Grand Unification is that the true
gauge {\bf group} (according to O'Raifeartaigh \cite{OR} one can to some
extend assign a meaning to a gauge Lie Group and not only to Lie algebra as
usually, except for on a lattice)  in our model is only, what we call the
Standard Model {\bf
  group} 
$S(U(2)\times U(3))\subset SU(5)$. So SU(5) is broken from the very beginning.
The idea to explain the factor 3 we had to put in
should be, that there in Nature indeed is one lattice for each family of
fermions in the Standard Model, meaning 3 layers of lattices. But the important
point for the series of energy scales of the present article is, that with this
interpretation of the deviation from SU(5) as a quantum correction (with somehow
the extra factor 3 due to the layers) allows us to avoid supersymmetry and
leads to a unification scale $5.3*10^{13}GeV$, very much (five and a half
order of
magnitudes) below the Plack scale,
$2*10^{19}GeV$, so that several different energy scales are indeed needed!

An energy scale already proposed in the article with the approximate
SU(5) unification is a scale, we call ``fermion tip'' scale, and which
is the scale obtained by putting the fermions in the Standard Model in a number
order after the mass, starting the enumeration at the top mass and going
downward in mass (counting the number of Weyl fermions, say, so that 
top counts for 6 and $\tau$ counts for two) and extrapolating to the number 0
in this counting. We found that there is a good fit with a parabola
touching the abscissa at the fermion number 0 at this ``tip of the
fermion spectrum'' being an extrapolated mass $10^4GeV$.

\subsection{Main Philosophy}

The main point of our work is to assume that we have a lattice
- this shall then be fluctuating in tightness, being somewhere tight,
somewhere rough with big links and net holes - and then the various
physical energy scales are calculated each of them from some power of the
length $a$ of a link. While for a rather narrow distrbution of a variable
$a$ say it is so that whatever the power of the variable $a$ you need for your
purpose you get about the same value for the effective typical $a$ size,
\begin{eqnarray}
  \sqrt[n]{<a^n>} &\approx& \hbox{ $n$-independent} \hbox{ for narrow
    distributions.}\\
  \hbox{However Galton distribtion: } P(\ln a)d\ln(a) &=&
  \frac{1}{\sqrt{2\pi \sigma}}\exp(-\frac{(\ln a - \ln a_0)^2}{2\sigma}) d\ln a
  \label{Galton}\\
  \hbox{gives rather }\sqrt[n]{<a^n>} &=& a_0\exp(\frac{n}{2} *\sigma).
  \label{ese}
\end{eqnarray}

So for a large spread $\sigma$ in the logarithm different typical
sizes $\sqrt[n]{<a^n>}$ become very different!
Imagining that the gravitational Einstein metric $g_{\mu\nu}$ is quantum
fluctuating
relative to some fundamental lattice, we take it to mean that the
lattice fluctuate relative to the metric in a way making its link size
flutuate even strongly in size relative to the metric tensor field.
As an ansatz we take the link size to have the Galton
distribution\cite{Kalecki,
  Pestieau,Johnson}
or let us just call it the Log Normal distribution. If a distribution
is composed of many {\bf multiplicative} fluctuating effects it will
by essentially the central limit theorem obtain this
Log Normal distribution
as pointed out by Gibrat\cite{Kalecki, Pestieau, Johnson}.
If the spread $\sigma$ of this Log Normal distribtion is large the different
$\sqrt[n]{<a^n>}$ can be of different orders of magnitude i.e. quite different.

To get an idea of how we may derive the relevant average of a power $<a^n>$
let us
for example think of a particle, a string, or a domaine wall being
described by action of the Nambu-Goto types
\begin{subequations}
\begin{eqnarray}
  \hbox{Particle action} S_{particle} &=&
  C_{particle}\int \sqrt{\frac{dX^{\mu}}{d\tau}g_{\mu\nu}\frac{dX^{\nu}}{d\tau}}
  d\tau\\
  &=& C_{particle}\int \sqrt{\dot{X}^2}d\tau\label{particle} \\
  \hbox{String action }S_{string} &=& C_{string}\int d^2\Sigma \sqrt{(\dot{X}\cdot
    X')^2 - (\dot{X})^2(X')^2}\\
  &=& -\frac{1}{2\pi \alpha'}\int d^2\Sigma \sqrt{(\dot{X}\cdot X')^2-
    (\dot{X})^2(X')^2}\label{string}\\
  \hbox{Domain wall action } S_{wall}&=& C_{wall}\int d^3\Sigma \\
  &&\sqrt{
    \det{  \begin{bmatrix} (\dot{X})^2 & \dot{X}\cdot X'&
      \dot{X}\cdot X^{(2)}\\
      X'\cdot \dot{X}& (X')^2 & X'\cdot X^{(2)}\\
      X^{(2)}\cdot \dot{X} & X^{(2)}\cdot X' & (X^{(2)})^2
    \end{bmatrix}}
    }\label{wall}
  \end{eqnarray}
\end{subequations}
Here of course these three extended structures are described by
repectively 1, 2,
and 3 of the parameters say $\tau, \sigma, \beta$, the derivatives with
respect to which are denoted by respectively $\dot{ }, ', and {}^{(2)}$. So
e.g. $d^3\Sigma = d\tau d\sigma d\beta$ and
\begin{eqnarray}
  X^{(2)}&=& \frac{\partial X^{\mu}}{\partial \beta}\\
  X'&=&\frac{\partial X^{\mu}}{\partial \sigma}\\
  \dot{X} &=& \frac{\partial X^{\mu}}{\partial \tau}\\
  \end{eqnarray}
Finally of course $\cdot$ is the Minkowski space inner product.

Now imagine, that in the world with the ontological lattice, which
we even like to take fluctuating, these tracks of objects, the
particle track, the string track or the wall-track, should be identified
with selections of in the praticle case a series of links, in the string case
a suface of plaquettes, and in the wall-case a three dimensional structure
of cubes, say. One must imagine that there is some dynamical marking of the
lattice objects - plaquettes in the string case e.g.- being in  an extended
object.  Now the idea is that we assume the action
for the lattice to have 
parameters
of order unity.
In that case the order of magnitude of the effective tensions meaning the
coefficients $C_{particle}, C_{string}, C_{wall}$ can be estimated in terms of
the statitical distribution of the link length - for which we can then
as the ansatz in the model take the Galton distribution (\ref{Galton}) -
by using respectively the averages of the powers 1, 2, 3, for our three
types of extended objects. I.e. indeed we say that by order of magnitude,
the mass of the particle, the square root of the string energy density
 or the string tension, and the cubic root of the domain wall tension
are given as the inverses of averages of $a$ like:
\begin{eqnarray}
  \hbox{Particle mass } m &\sim & <a>^{\; -1}\label{Pm}\\
  \hbox{Square root of string tension } \sqrt{\frac{1}{2\pi \alpha'}} &\sim &
  \sqrt{<a^2>}^{\; -1}\label{Stsr}\\
  \hbox{String tension itself } \frac{1}{2\pi \alpha'} &\sim& <a^2>^{-1}\\ 
  \hbox{Cubic root of wall tension } S^{1/3} &\sim& \sqrt[3]{<a^3>}^{\; -1}
  \label{Wtcr}\\
  \hbox{wall tension itself } S &\sim& <a^3>^{-1}.
  \end{eqnarray}
Here the $\sim$ approximate equalities are supposed to hold order of
magnitudewise under the assumption that no very small or very big numbers are
present in the coupling parameters of the lattice, so that it is the somehow
averaged lattice that gives the order of magnitude for these energy densities
or tensions.

\subsection{Illustration of the idea}
Although our speculations for the three energy scales - meaning numbers with
dimension of energy - which we in my speculation attach to these three
objects, the particle, the string, and the wall, are indeed very speculative
only,
and that we shall give a bit better set of such scales in next subsection, let
us nevertheless as a pedagogical example consider these three first:

{\bf The speculative pedagogical example:}
The three speculative scales are chosen as:

\begin{itemize}
\item{\bf The particle} Remembering that on a lattice with gauge particle
  degrees of freedom you may due to the very lattice get monopoles, the time
  tracks of which are described by series of cubes in the lattice, we propose
  to identify the ``particle'' described by an action of the type
  (\ref{particle}) with a monopole caused by the lattice structure for the
  lattice gauge theory of the Standard Model.

  One of the very few new physics particles perhaps found in the LHC
  and even by studies of the data from LEP is a dimuon resonace with
  mass 27 GeV found in data selected with some b-meson activity\cite{dimuon,
    Heister}. We speculate that this hopefully existing dimuon decaying
  particle is related to some monopoles for the Standard Model group -
  possibly a bound state by some sort of confinement of a couple of
  monopoles - with monopole mass of the order
  \begin{eqnarray}
    \hbox{``monopole mass''} &\sim& 27 GeV\label{monopole}.
    \end{eqnarray}
  
  Taking it that all other parameters than the ones in the link
  distribution are of order unity, we get
  using (\ref{ese}, \ref{Pm}) up to order unity factors:
  \begin{eqnarray}
    27 GeV &\approx& <a>^{-1} =a_0^{-1}\exp(-\frac{\sigma}{2}).\label{rparticle} 
    \end{eqnarray}
  
\item{\bf String} We shall identify the strings with the {\bf hadronic
  strings}, if we dreamt of some realistic type of fundamental superstring,
  we would have no chanse to fit our fluctuating lattice with the other
  proposal we are making. So we take it that the moderate success of
  Veneziano models and string models\cite{histstring} in hadron physics
  means, that the hadrons
  are at least in some crude approximation strings. This is true because
  confinement means, that a quark and an antiquark will be held together
  by a potentially long gluon structure, which functions like a string.
  In the string model for the hadrons the string tension or equivalently
  their energy density are given as $\frac{1}{2\pi \alpha'}$ where the
  $\alpha'$ is the Regge slope meaning the derivative of the angular momentum
  versus the square of the mass(this slope seems to vary rather little along
  the Regge trajectory, as well as from trajectory to trajectory). In our
  old paper\cite{slope} Ninomiya and I used the empirical value $\alpha'
  =0.88 GeV^{-2}$ and thus
  \begin{eqnarray}
    \hbox{Energy density } \frac{1}{2\pi \alpha'} &=&
    \frac{1}{2 \pi 0.88GeV^{-2}}\\
    &=& 0.181 GeV^2\\
    \hbox{or energy scale } \sqrt{\frac{1}{2\pi\alpha'}}&=&0.43 GeV.
    \label{alphaprim}\\
    \hbox{Also we thought of } \hbox{``Hagedorn temperature''} &=& 0.16 GeV
    \label{Hagedorn}
    \end{eqnarray}

  So we get assuming other parameters than the ones in in link distribution
  being of order unity and using (\ref{ese},\ref{Stsr}) up to order unity
  factors:
  \begin{eqnarray}
    0.43 GeV &\approx & \sqrt{<a^2>}^{\; -1} = a_0^{-1}\exp(- 2\frac{\sigma}{2}).
    \label{rstring}
  \end{eqnarray}
  
\item{\bf Domaine wall} Colin Froggatt and myself have for long developped
  a model for dark matter \cite{odm}, in which there are (at least) two
  different
  phases of the vacuum and domaine walls in between them, where these phases
  meet each other. Fitting various informations on the dark matter
  we arrived in our latest work \cite{odm} in table 2 the sopposedly best
  values, 12 MeV and 4 MeV for the cubic root of the tension of the domaine
  wall. It was  found
  by two different methods in line 2. and 3. respectively. The value 12 MeV
  were derived using the rate of the DAMA observations, while the value
  4 MeV was based on the hypotesis, that the stopping length of our dark matter
  pearls, when hitting the Earth, were just so as to make  the pearls stop
  at the
  depth 1400m of the DAMA experiment, favouring that just this experiment DAMA
  sees the dark matter, while no other ones see it. In our model radiation
  of the dark matter with its 3.5 keV radiation, which is with
  trouble seen astronomically\cite{Xr35}, is supposed to be what underground experiments
  potentially see, but only DAMA saw it. The deviation between the two estimates
  of the cubic root of the wall, 12 MeV and 4 MeV, as only a factor 3 should
  be considered a sign of consistency or success of our model and a
  geometrically averaged cubic root of the tension,
  7 MeV could be taken.
  \begin{eqnarray}
    \hbox{cubic root of tension} &=& 7 MeV= 0.007 GeV\label{3root}\\
    \hbox{tension}&=& (7MeV)^3= 2*10^2 MeV^3
  \end{eqnarray}
  
  Thus modulo order unity factors our fluctuating lattice here using
  (\ref{ese}, \ref{Wtcr}) gives:
  \begin{eqnarray}
    7 MeV &\approx& \sqrt[3]{<a^3>}^{\; -1} = a_0^{-1}\exp(-\frac{\sigma}{2}*3).
    \label{rWall}
    \end{eqnarray}
  
  \end{itemize}

The check of our model - the fluctuating lattice with the Galton link size
distribution (\ref{Galton}) - means for these three energy scales that the
following two
ratios are equal to each other:
\begin{eqnarray}
  \exp(\frac{\sigma}{2}) &=& \frac{27 GeV}{0.43 GeV}=63
  \hbox{(from ``particle'' and ``string'')}\\
  \exp(\frac{\sigma}{2})&=& \frac{0.43GeV}{7 MeV}=61
  \hbox{(from ``string'' and  ``wall'')}\\
  \hbox{Success with } \sigma &=& 8.3\label{s83}\\
  a_0^{-1} &=& 1.7 *10^3 GeV.\label{a0m117}
\end{eqnarray}

It is indeed very strange, if the energy scale of the approximative string
theory describing the hadron physics were connected with a fundamental
lattice - such as our fluctuating one - because we know from QCD, that the
scale of these hadronic strings is gives by the running strong Yang Mill
coupling and not from any lattice a priori.

\section{Our main set of Energy Scales}
At the time of the talk in Corfu I had not yet thought upon the just presented
three energy scales, but was rather occupied the {\bf approximate SU(5)
  unification scale}, the (reduced) Planck scale, and a scale, which I invented
myself called the {\bf ``fermion tip'' scale}:

\begin{itemize}
\item{\bf Fermion tip scale} In principle you can count all the fermions
  in the standard model after their mass - except that some are degenerate
  due to color symmetry - and then you can fit the number of the fermion
  counted from the highest mass downward versus the logarithm of the mass
  with a smooth curve. We claim that doing that a parabola like curve
  tounching the number zero at the mass $10^4 GeV$ fits well. In this sense
  we can claim that the ``tip of the fermion spectrum'' is at the
  ``extrapolated mass'' $10^4 GeV$. In the tables we have for each bunch of
  degenarate fermion states (counted as number of Weyl fermions) put the
  average number in the series counted from above and the logarithm
  (with basis 10 for simplicity) and from there constructed a number that
  should be constant in the sense of being the same for all the bunches of
  fermions
  provided the fermion number in the series indeed fitted the just mentioned
  parabola. That a curved fitting curve is favourable to a line fit
  of the number in the series versus the logarithm of the mass is seen
  from the figure \ref{ft}. Our tables shown give for each bunch
  of fermions in the Standard model a construction of a number, that will
  be constant provided the fermions ordered this way indeed lies on
  the parabola, and indeed it is seen to be about 1.1 to 1.2 mostly.

  We want to assume, that in the region of masses below the
  tip point - which we call $m_{mnl}$ -  the number of fermions
  which can still at a certain scale be considered massless shall be
  proportional to the number of links the inverse size of which equals
  that scale. With such philosophy - to be explained more in the articles we
  review \cite{Universe} - the tip point $m_{mnl}$ should be the
  maximal probality point for the (logarithmic) distribution of the
  inverse link size. I.e. we should take
  \begin{eqnarray}
    m_{mnl} &=& 1/a_0.
    \end{eqnarray}

  \begin{eqnarray}
    {\bf Quark\; for\; m_{mnl}}&=&{\bf 10^4GeV }\label{mnmlq}
    \end{eqnarray}
 
\begin{adjustbox}{width=\textwidth, center}
\begin{tabular}{|c|c|c|c|c|c|c|}
  \hline
  Name
  & number $n$&Mass $m$&$log_{10\;
    GeV
    }m$&$diff$ = $4\text{-}log m$&
  $diff^2$&$diff^2/n$\\
  \hline
  top&3~$\pm$~1&172.76  ~$\pm$~0.3  GeV&2.2374~$\pm$~0.0008&1.7626&3.1066~$\pm$~0.003&1.0355
  ~$\pm$~0.001 $\pm $0.4\\
  \hline
  bottom&9~$\pm$~0.3& 4.18~$\pm$~0.0079  GeV& 0.6212~$\pm$~0.001&
  3.3788&11.416~$\pm$~0.01&1.268~$\pm$~0.001~$\pm$~0.03 \\
  \hline
  charm&17 or 15&1.27~$\pm$~0.02&0.10382~$\pm$~0.009&3.8962&15.180~$\pm$~0.07&
  0.893 ~$\pm$~0.004 ~$\pm$~0.06\\
  \hline
  strange&25 or 23&0.095~$\pm$~0.006  GeV& $-$1.0223~$\pm$~0.003 &5.0223&25.223
  ~$\pm$~0.03&1.009~$\pm$~0.001~$\pm$~0.1\\
  \hline
   down& 31&4.79~$\pm$~0.16 MeV&$-$2.3197~$\pm$~0.01 &6.3197&39.939~$\pm$~0.06&1.288
   ~$\pm$~0.002 \\
   \hline
  up&37&2.01~$\pm$~0.14 MeV&$-$2.6968~$\pm$~0.03& 6.6968&44.847~$\pm$~0.4&1.212~$\pm$~
  0.01\\
  \hline
\end{tabular}
\end{adjustbox}
 %
  \begin{eqnarray}
    {\bf Leptons\; for\; m_{mnl}}&=&{\bf 10^4GeV }\label{mnmll}
    \end{eqnarray}
\begin{adjustbox}{width=\textwidth, center}
\begin{tabular}{|c|c|c|c|c|c|c|}
  \hline
  Name& number $n$&Mass $m$&$log_{10\;
    GeV
  }m$ 
  &   
$diff$ = $4\text{-}log m$&
  $diff^2$&$diff^2/n$\\
  \hline
  $\tau$&13 or 19&1.77686 $\pm$ 0.00012&0.2496 $\pm$  0.00003&3.7503&14.065 $\pm$ 
  0.0003&1.082
   $\pm$  0.00002 $\pm $0.4\\
  \hline
  mu&21 or 27&105.6583745 $\pm$  $2.4*10^{-6}$~MeV&$-0.9761
  \pm$ $10^{-8}$   
&4.9761&
  24.761 $\pm$  $10^{-7}$&1.179 $\pm$  $4*10^{-9}$  $\pm$  0.3\\
  \hline
  electron&41&0.51099895069 $\pm$ $1.6*10^{-10}$& $-$3.2915 $\pm$  $4*10^{-10}$&
  7.2916&53.167 $\pm$  $10^{-8}$&1.297 $\pm$  $10^{-11}$\\
  \hline
\end{tabular}
\end{adjustbox}

\begin{figure}[h]
  \includegraphics[scale=0.6]{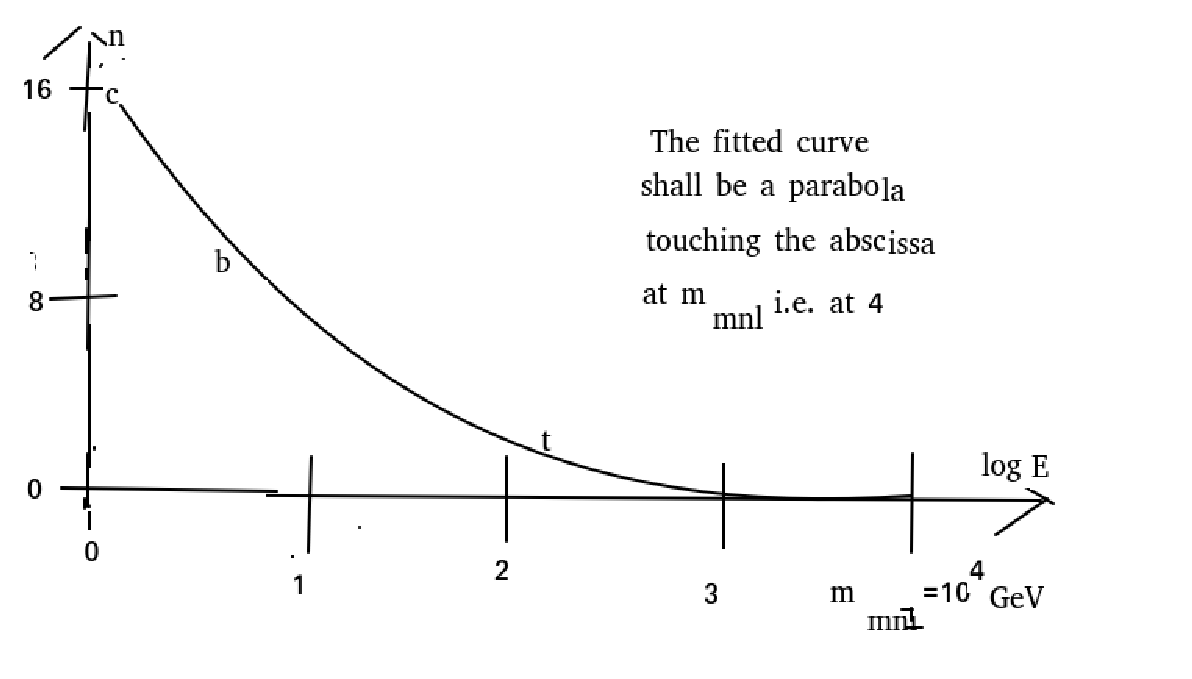}
\caption{ \label{ft}
  Because
 the distribtuion of log links after size has a
  maximum
  at the fermion tip point, we suggest that also the density of activ
  fermions (=fermions with lower mass than the scale $E$, at which you ask
  for the density of links) should behave this way, meaning with a
  {\bf parabola} behavior with maximum at the point $m_{nml}$ (=the fermion
  tip point).
  If we take it that there are 45 or 44 chiral fermions per family the number
  of  families will at chiral fermion number $n$ counted from the top
  downwards in mass, be
  {``Number of active families''} = (3 $*$ 45 $-$ n)/45.
  Note that the three shown point, c, b, and t, fit better on the
  parobola than on a straight line, and so the fitting form chosen is
   somewhat empirically supported.
}
\end{figure}

This ``Fermion tip'' scale is to be thought of as the analogue to the other
scales  with their form $\sqrt[n]{<a^n>}$ but with $n=0$. However $n=0$
does not quite make a good sense in as far as it would only give
$\sqrt[0]{1}$, which is nonsense. So instead we must take the best we can:
\begin{eqnarray}
  \hbox{``Fermion tip''-scale} &=& \exp(<\ln(a)>)\\
  \hbox{formally n=0}&&\nonumber
\end{eqnarray}

\item{\bf Approximate unification scale} This is the scale at which the
  running (inverse) fine structure constants  meet in the ratios predicted
  by grand unfied SU(5). It is wellknown that without susy\cite{susy} or
  some similar
  modification the  running fine structure constants do not meet as they
  should in grand unified SU(5); but there is a little interval of energy
  scales in which $U(1)$ fine structure constant (corrected with its
  3/5 for SU(5) notation) lies in between the two non-abelian fine
  structure constants, and one can say that in this region the three
  fine structure constants are more concentrated than at other
  energy scales. So we can take this rather small interval as
  ``the approximate unification scale''. In \cite{AppSU5} we propose that
  with a 3 times magnified quantum corrections to a classically SU(5)
  symmetric action we get after the tripled quantum correction understanding
  of the running couplings as such a quantum corrected (with 3 times too much)
  classical SU(5) set of couplings. This gives a rather accurate unification
  inside the just mentioned interval of scales.

  Now as is easily seen the effective Lagrangian density for the gauge theories
  evaluated in a lattice theory is given by the coulings on the single
  plaquattes multiplied by a number of the order $a^{-4}$ due to that
  the Langian density per $(length\;  unit)^4$ is of course proportional
  to the number of hyper-cubes in such a 4-volume. This number is
  scaled up of course by the $a^{-4}$. The very crossing of the running
  couplings has character of a logarithmic relation, and we believe it
  will not change how the unification scale is to be extracted from the
  link-distribution. So we take it:
  \begin{eqnarray}
    \hbox{``Approximate unification'' scale } m_U &\approx& \sqrt[-4]{<a^{-4}>}\\
    &=& \frac{1}{\sqrt[4]{<a^{-4}>}}.\\
      \hbox{I.e. } n&=& -4.
    \end{eqnarray}

  Let us evaluate this value $n=-4$ (again)  by noticing the way the various
  factors in
  the action for a Yang Mills theory -or just electro dynamics - behave as
  functions of the link length $a$ when one latticize this Yang Mills theory:
  \begin{eqnarray}
    \hbox{Action } S &=& \int \frac{-1}{4e^2}F_{\mu\nu}F^{\mu\nu}d^4x\\
    \hbox{or with metric } S &=& \int \frac{-1}{4e^2}  g^{\mu\rho}g^{\nu\tau}
    F_{\mu\nu}F_{\rho\tau}\sqrt{g}d^4x\\
    \hbox{where say for EM: } F_{\mu\nu}&=& \partial_{\mu}A_{\nu}-
    \partial_{\nu}A_{\mu}
    \end{eqnarray}

  In ``latticising'' we get the $a$ dependences:
  \begin{eqnarray}
    \hbox{\bf Without metric: }&&\nonumber\\
    \int ...d^4x & \rightarrow& <a^4>\Sigma_{sites} ...\\
    \partial_{\mu} &\propto& a^{-1}\\
    A_{\mu} &\propto& a^{-1}\\
    \hbox{``Space-time-volume''} V_4 &=& \int d^4x \rightarrow
    <a^4>\Sigma_{sites}1\\
    \hbox{So action } S &\propto&``integral \; of'' a^{-4}<a^4>\\
     \hbox{so } \frac{S}{V_4} &\propto& <a^{-4}>\\
    \hbox{while {\bf With metric}}&&\nonumber\\
    \int ...d^4x &\rightarrow&<a^4>\Sigma_{sites}\frac{1}{\sqrt{g}}...\\
    \hbox{or } \int ...\sqrt{g}d^4x &\rightarrow&<a^4> \Sigma_{sites} ...\\
    g_{\mu\nu} &\propto& a^2\\
    g^{\mu\nu} &\propto& a^{-2}\\
    \partial_{\mu} &\propto& 1\\
    A_{\mu} & \propto& 1\\
     \hbox{``Space-time-volume''} V_4 &=& \int\sqrt{g} d^4x \rightarrow
    <a^4>\Sigma_{sites}1\\
    \hbox{So that } S = \int g^{\mu\nu}g^{\rho\tau}F_{\mu\rho}F_{\nu\tau}\sqrt{g}d^4x
    &\propto& ``integral \; of''a^{-4}<a^4>\\
    \hbox{and thus }\frac{S}{V_4}& \propto& <a^{-4}>
  \end{eqnarray}

  In our article \cite{AppSU5} we found by our model of SU(5) symmetry
  being exact in a lattice theory in the classical approximation
  but being corrected by a quantum correction 3 times the
  a priori true one that the unification scale for this only classicially
  SU(5) unification was
  \begin{eqnarray}
    m_U &=& 5.3*10^{13}GeV,\label{unification}
  \end{eqnarray}
  which is matching with just looking for the scale, at which the running
  couplings
  in the minimal SU(5) notation are closets to each other.

  (Had we believed in susy the unification energy scale would
  have been rather
  \begin{eqnarray}
    \hbox{susy unification}&=&10^{16} GeV,\label{susy} 
    \end{eqnarray}
  but that would {\em not} fit well in our scheme. So our phenomenolgical
  line fit, disfavours susy!)
  
\item{\bf Planck energy scale}
  The Einstein field equation looks
  \begin{eqnarray}
    R_{\mu\nu}-\frac{1}{2} R g_{\mu\nu} &=&\kappa T_{\mu\nu}\\
    \hbox{or in trace reversed form } R_{\mu\nu} &=&\kappa
    (T_{\mu\nu}-\frac{1}{2}Tg_{\mu\nu}),
    \end{eqnarray}
  where then the left hand side is the curvature while the right
  hand side represent the energy momentum tensor.
  Thinking of a single hyper cube in the lattice the curvatur is to be
  represented as a discreticed second derivative of some variables
  defined on the lattice. Therefore on a single hypercube the
  ratio
  \begin{eqnarray}
    \hbox{second derivative at hypercube} &\propto& a^{-2}\\
    \hbox{number hypercubes per four-volume} & \propto& a^{-4}\\
    \hbox{so naively  } \frac{1}{\kappa} \propto a^{-2}.
  \end{eqnarray}
  But this was naive because we should have the second derivative and the
  energy or momentum sitting on the {\bf same} hypercube so that the special
  gravitational interaction between them comes in.
  When the $a$ is fluctuating and one shall average over,
the common place for a second derivative and a piece of energy momentum 
This would mean
  inside a unit length hypercube the second derivative providing
  hypercube should be one among of the order of $a^{-4}$ ones. Such a restricted
  second derivative would be a factor $a^{-4}$ lower in probability of being
  found. So:
  \begin{subequations}
  \begin{eqnarray}
    \hbox{ Restricted } \partial^2 = (\partial_{\mu}\partial_{nu})_{at \; given\;
      hypercube} &\propto& a^{-2}/a^{-4}/\hbox{``unit four space time vloume''}\\
   & =& a^2/\hbox{``unit 4-volume''}\\
    \hbox{Taking it }g_{\mu\nu} && \hbox{ independent of $a$}\\
    \hbox{and Einstein equation of form:} \partial^2 ``{g_{\mu\nu}}'s'' &\propto&
    \kappa  a^{-4}*``{g_{\mu\nu}}'s''\\    
    \hbox{or } \partial^2 &\propto&\kappa a^{-4}\hbox{(mod $g_{\mu\nu}$
      factors, allowing diemensions different)} \\
      \hbox{so that with } \partial^2 &= & \hbox{Restricted } \partial^2
      \propto a^2/\hbox{``unit 4-volume''}\\
      \hbox{we get } \frac{1}{\kappa} &\propto& a^{-6}*\hbox{``unit 4-volume''}
      \\
      \hbox{If relevant power is for } \frac{1}{\sqrt{\kappa}} &is & a^{-6}\\
      \hbox{and we just use dimensionality } \frac{1}{\sqrt{\kappa}} & \sim&
      \sqrt[6]{<a^{-6}>} = a_0\exp(\frac{\sigma}{2}*6);
  \end{eqnarray}
  \end{subequations}
  but this last step sounded dangerous.
  
  This should mean that order of magnitude of the coefficient $\kappa$ in the
  Einstein field equation, provided the lattice parameters are of order
  unity, should be given by averaging a sixth power so that we expect
  \begin{eqnarray}
    \hbox{Reduced Planck constant} \frac{1}{\sqrt{\kappa}}
    &\approx & \sqrt[6]{<a^{-6}>}
  \end{eqnarray}
  (This calculations of the relevant powers deserves some further development,
  and a  slightly different calculation is found in \cite{Universe}, and is
  actually alluded to in the table.)

  In the notation with $\hbar = c=1$
  \begin{eqnarray}
    \kappa = 8\pi G & =& 2.07665*10^{-43}N\\
    &=& 2.07665*10^{-43}N*8.19*10^5N/GeV^2\\
    &=& 1.7008*10^{-37}GeV^{-2}\\
    \hbox{so that } \hbox{``Reduced Planck energy''}=\frac{1}{\sqrt{\kappa}}
      &=& \frac{1}{1.7008*10^{-37}GeV^2}\\
      &=& (4.124*10^{-19}GeV)^{-1}\\
      &=& 2.425*10^{18}GeV.\label{ReducedPlanck}
  \end{eqnarray}

  With ignoring the $8\pi$ we get the unreduced Planck scale
  \begin{eqnarray}
    \hbox{Planck scale} \hbox{(not reduced)} &=& 1.22*10^{19}GeV
    \label{Planckenergy}.
    \end{eqnarray}
\end{itemize}

Now fitting to (\ref{ese}) we have
\begin{eqnarray}
  \frac{\hbox{``Reduced Planck scale''}}{\hbox{``Unification scale''}}
    &=& \frac{2.424*10^{18}GeV}{5.3*10^{13}GeV}\\
    &=& 4.58 *10^4\\
    \hbox{so } \exp(\sigma/2) &=& \sqrt{4.58*10^4}\\
  &=& 214; \\
  \hbox{to be compared with: } &&\\
  \frac{\hbox{``Unification scale''}}{\hbox{``Fermion tip scale''}}&=&
    \frac{5.3*10^{13}GeV}{10^4 GeV}\\
    &=& 5.3*10^9\\
    \hbox{so that } (\exp(\sigma/2))^4 &=& 5.3*10^9\\
    \hbox{giving } \exp(\sigma/2) &=& \sqrt[4]{5.3*10^9}\\
    &=& 270
  \end{eqnarray}
Order-of-magniudewise at least 214 and 270 are very close, so the fit to the
Galton distribution is good.

If we accept that the at first gotten 61 to 63 are also of the order
of 214 to 270, then we have now  about 6 different energy scales fitting
on the straight line of the logarithm of the energy versus the power $n$
to which the link size $a$ should be taken before the average and the
for dimensionality needed $n$th root of the average is taken. We had namely
approximately same value for the $a_0^{-1}$, namely $1.7 *10^3 GeV$ versus
$10^4 GeV$.(See (\ref{a0m117}, \ref{mnmlq}, \ref{mnmll})).

Indeed we have invented/found in litterature  a couple or three more such
energy scales, believe it or not, we can claim they fit also into the same line
in the logarithm versus $n$ plot, so that we now have about 8 to 9 such
energy scales. When I give the number we reached to as 8 to 9 it is
because we can count the two scales connected with inflation of the universe,
``the infation time Hubble-Lemaitre expansion rate $H$'' and the
``fourth root of the inflaton potential $V_{inflation}$
i.e. $\sqrt[4]{V_{inflation}}$'' as two different ones or as essentially the
same.(The information used for these cosmological energy scales was extracted
from \cite{anintr, Enquist, Bellomoetal}. )

The problem with many of the extra scales found is that they are typically
like the ``scale of the see-saw neutrino masses'' not only dependent on the
phenomenologically accessible information, as for the see-saw scale
the neutrino oscillation data, but on some theoretically guessed model.
This makes the true uncertainty of this see-saw-scale several orders
of magnitude, but the value, which come from most detailed theoretical
models of ``See saw mass scale'' = $10^{11}GeV$, fits wonderfully our
straight line from the Galton distribtuion.

\section{The Full Table}

For completeness we include the table of all the 8 to 9 energy scales
from \cite{Universe}, and to follow the integer number $q$ in the reference
is related to our power $n$ to which the link size is raised when the
average is taken by
\begin{eqnarray}
  q&=& n +4.
  \end{eqnarray}

\subsection{Short Explanation of the table}
For most details of the table now to be given we refer to the
long article \cite{Universe} or the description after the table;
but shortly you should notice the symbols of the form
$a^n$ in the second column and the third line inside each of the blocks;
this is the expression  - a power, the $n$th power, of the link length
$a$ - over which the average under the lattice fluctuation - with the
Galton distribution - is to be taken, when evaluating the (our) lattice theory
prediction of the order of magnitude of the energy scale, whos name
is put in the first line inside its block in the first column. Of course for
dimenional reasons the energy scale must then be
\begin{eqnarray}
  ``the\;  energy\; scale'' &\approx& \sqrt[n]{<a^n>}^{\; -1}\\
  \hbox{when we use } a^n.&&
\end{eqnarray}

The second thing to note in the table are the two numbers in line
1 and 2 in the blocks in column 3. The uppermost number inside the
block is the phenomenologically or experimentally given number for the
energy scale, while the number below is the prediction of our fit
- fitting the numbers $1/a_0$ and $\exp(\sigma/2)$ - for the energy scale
energy in question. It is thus the order of magnitude agreement of these
two numbers in column number 3, that express the success of our model
for the energy scale of the block in question.

For several of the energy scales there are small variations in how the
exact definiton of the scale should be taken or the like, and therefore
some scales have gotten a couple of blocks, then seprated by a single
horizonatal line, while one energy scale and the next are seperated by a
double horizontal line.

\subsection{Short mention of those Scales, we Essentially Skipped in this
  article}

After looking at the articles\cite{King,Davidson, Grimus, Mohapatra,
  Takanishi, Takanishi2}
we decided the guess

\begin{eqnarray}
  \hbox{``see-saw scale''} &\approx& 10^{11} GeV,\label{seesaw}
  \end{eqnarray}

and similarly after looking at \cite{Enquist, anintr, Bellomoetal}
we decided to take for the situation at (end of) infaltion: 
\begin{eqnarray}
  \hbox{``Infl. Hubble const.'' } H_{infl.} & \approx& 10^{14} GeV
  \label{inflationH}\\
  \sqrt[4]{\hbox{``Infl. potential''}} V_{infl.}^{1/4} &\approx& 10^{16} GeV
  \label{inflationV}
\end{eqnarray}

Concerning these cosmological scales about inflation, I do not feel safe
as to what should be our power $n$, but believe we got it right with
$n=-5$ for the $V_{infl.}^{1/4}$ and $n=-4$ for the $H_{infl.}$.
(The information on the energies were extracted from
\cite{anintr,Enquist,Bellomoetal}).

Finally since the ``scalars scale'' is only my phantasy, there is no direct
observation of it; but the speculation is that there exist with the mass of
this ``scalars scale'' a lot of scalar bosons, some of which have
vacuum expectation values breaking spontaneously some symmetries of importance
for making the masses of Standard Model Fermions having big mass rations
(so called small hierarchy problem), and since the fermions from the
``see-saw-scale'' are supposed to be involved as intermediate fermions
in the masses for the Standard Model ones, the ratio of the
of the ``see-saw scale'' to the ``scalars scale'' provides the small number
making the breakings of the symmetries small. From this kind of model for
weak breaking of symmetries and thus large mass ratios among the Standard
Model 
Fermions we expect such ratios to be of the order of
$\frac{``scalars\;  scale''}{``see-saw \; scale''} =251$ from our
straight line, to be compared with a typical ratio of Standard Model
Fermions
43 (
or corrected by a $\sqrt{4\pi}$ rather 152.)

\subsection{The very table}

\begin{adjustbox}{width=\textwidth}

  \begin{tabular}{|c|c|c|c|
    }
  \hline
  Name  comming from status
&[Coefficient] Eff. $q$ in $m^q$ term&``Measured'' value Our Fitted value&Text ref. Lagangian dens.  \\
  \hline
  \hline
  Planck scale&$[mass^{2}]$ in
  kin.t.
  &$1.22*10^{19}$~
  GeV
  &(\ref{Planckenergy}) \\
  Gavitational $G$&
  q
  ~=~$-$2  
  &$2.44*10^{18}$~
  GeV
  &$\frac{R}{2\kappa} $ \\
  wellknown&$a^{-6}$&&$\kappa~=~ 8\pi G $ \\
  \hline
  Redused Planck&$[mass^{2}]$ in
  kin.t.
  &$2.43*10^{18}$~
  GeV
  &(\ref{ReducedPlanck}) \\
  Gravitational $8\pi G$&
  q
  ~=~$-$2&$2.44*10^{18}$~
  GeV
  &$\frac{R}{2\kappa}$ \\
  wellknown&$a^{-6}$&&$\kappa~=~8\pi G$ \\
  \hline\hline
  Minimal $SU(5)$&
  $[1]$
  &$5.3*10^{13}$~
  GeV
  &(\ref{unification}) \\
  fine structure
  const.s
  $\alpha_i$&
  q
  ~=~0&$3.91*10^{13}$~
  GeV
  &
  $\frac{F^2}{16\pi \alpha}$ \\
  only approximate&$a^{-4}$&&$F_{\mu\nu}~=~\partial_{\mu}A_{\nu}-\partial_{\nu}A_{\mu}$ \\
  \hline
  Susy $SU(5)$&
  $[1]$
  &$10^{16}$~
  GeV
  &(\ref{susy}) \\
  fine structure
  const.s
    &
    q
    ~=~0&$3.91*10^{13}$~
    GeV
    &$\frac{F^2}{16\pi \alpha}$ \\
  works&$a^{-4}$&&$F_{\mu\nu}~=~\partial_{\mu}A_{\nu}-\partial_{\nu}A_{\mu}$ \\
  \hline
  \hline
  Inflation $H$&
  $[1]$
  ?
  &$10^{14}$~
  GeV
  &(\ref{inflationH}) \\
  CMB, cosmology&
  q
  ~=~0  ?&$3.91*10^{13}$~
  GeV
  &$\lambda \phi^4$ \\
  ``typical'' number&$a^{-4}$&&$V~=~\lambda \phi^4$ \\
  \hline
  Inflation $V^{1/4}$ &concistence  ?&$10^{16}$~
  GeV
  &(\ref{inflationV})  \\
  CMB, cosmology&
  q
  ~=~$-$1  ?&
  $9.96*10^{15}$~
  GeV
  &consistency \\
  ``typical''&$a^{-5}$&&$V~=~\lambda \phi^4$~ ? \\
  \hline
  \hline
  See-saw
  & $[mass] ~in \; non\text{-}kin.$ &$10^{11} $~
  GeV
  & (\ref{seesaw}) \\
  Neutrino oscillations&
  q
  ~=~1& $1.56*10^{11}$~
  GeV
  &$m_R\bar{\psi}\psi$ \\
  modeldependent&$a^{-3}$ &&$m_R$ right hand mass \\
  \hline\hline
  Scalars&$[mass^2]~ in\; non\text{-}kin.$&$\frac{seesaw}{44\; to\; 560}$&(\ref{s43}) and  (\ref{s152}) \\
  small hierarchy&
  q
  ~=~2&$\frac{1.56*10^{11}~\text{GeV}}{250}$& $m_{sc}^2|\phi|^2$ \\
  invented by me&$a^{-2}$&&breaking $\frac{seesaw}{scalars}$ \\
  \hline\hline
  Fermion tip& $``[mass^4] ~in\; non\text{-}kin.\text{''}$& $10^4 $~
  GeV
  &
  (\ref{mnmll}) and (\ref{mnmlq}) \\
  fermion masses&
  q
  ~=~4& $10^4$~
  GeV
  & ``1'' \\
  extrapolation &$a^0=1$&&quadrat fit \\
  \hline\hline
  Monopole& $``[mass^5]~ in\; non\text{-}kin.\text{''}$& $28 $~
  GeV
  &(\ref{monopole}) \\
  dimuon 28 GeV&
  q
  ~=~5 & $40 $~
  GeV
  &$m_{monopol}\int ds$ \\
  invented&$a$&&$S \propto a$ \\
  \hline\hline
  String $1/\alpha'$ & $``[mass^6]~in\; non\text{-}kin.\text{''}$& $1 $~
  Gev
  & (\ref{alphaprim})  \\
  hadrons&
  q
  ~=~6& $0.16 $~
  GeV
  &Nambu Goto \\
  intriguing & $a^2$ &&$S\propto a^2$ \\
  \hline
  String $T_{hagedorn}$&
  ``
  $[mass^6]$  
    in\; non-kin.''
    & $0.170$~
    GeV
    & (\ref{Hagedorn}) \\
hadrons &
q
~=~6& $0.16 $~
GeV
& Nambu Goto \\
 intriguing  & $a^2$&& \\
 \hline\hline
 Dom. wall&$[mass^7]$ in non kin.& 0.007~GeV &(\ref{Wtcr}, \ref{3root})\\
 dark matter& q=7& 0.00064~GeV & Goto 3 dim. \\
 intriguing& $a^3$&&\\
 \hline
\end{tabular}
\end{adjustbox}

\subsubsection{Content in the Different Coloumns}

\begin{itemize}

\item{\bf
  The first column (from left)}
  contains the three items:   
\begin{itemize}
\item[1.]
  A name,
 we just acsribe to the scale in question.

\item[2.] An allusion to, from which data the energy scale number is determined.

\item[3.] What we call ``status'', an estimation of how good the story
  of the scale in question ~ is.
\end{itemize}

\item{In {\bf the second column} the items are:}
\begin{itemize}
\item[1.] $[coefficient]$: It is the dimensionality of the coefficient
  to a term
in the Lagrangian relevant for the interaction giving the scale in question.
In the table is added either ``
ìn
kin.t.
  '' or ``in non-kin.'', meaning
respectively, that the term with the coefficient of the dimension given
is the kinetic term or the non-kinetic term repectively.

\item[2.] In next line inside this column 2 is written the quantity $q$
  and its value
for the scale in question, or some effective value for this $q$.
For non-kin. $q=dim_{energy}(coefficient)$, while for ``
kin.t.
'' it is
$q=-dim_{energy}(coefficient)$ because we have in all of cases (by accident),
that the other
coefficient is dimensionless.

\item[3.] The power $n$ of the link size $a$ the average of which is supposed
  to give the size of the (coupling constant for) energy scale in question
  expressed by the symbol $a^n$. It means that the energy scale
  is given as
  \begin{eqnarray}
    \hbox{`` energy scale''} &=& \sqrt[n]{<a^n>}^{\; -1}= \frac{1}{a_0}
    \exp(-n \frac{\sigma}{2}).
    \end{eqnarray}
  
\end{itemize}

\item{Then in the {\bf third column}:} comes the numbers, the energy scale.
\begin{itemize}

\item[1.] In the top line inside the blocks comes the experimental or rather
   best theoretical estimate from the experimental data.(the type of data was
   mentioned
in second line in column 1).

\item[2.] In the next line we have put the fit to the straight line of the
  logarith of
  the energy versus $q=n+4$. It is given by the fitting formula:
  \begin{eqnarray}
    \hbox{``fitted value'' }&=& 10^4~\text{GeV}*250^{-n}\\
    &=& 10^4~\text{GeV}*(250)^{4-q}\\
    &=&3.91*10^{13}~\text{GeV}*250^{-q}\\
    &=&3.91*10^{13}~\text{GeV}*250^{-n+4}
    \end{eqnarray}
\end{itemize}

\item{In {\bf fourth column}:}
  \begin{itemize}
\item[1.] In first line is a reference to the formula in the text
    representing the decicion as to, what value to take for the scale in
    question. (Usually at best trustable to order-of~magnitude).
\item[2.] In the second line is the Lagrangian density used in determining
    the dimension of its coefficient, $[coefficiient]$.
\item[3.] In the third line we put some formula or remark supposed to
    make it recognizable, what the Lagrangian density in the line 2
    means.
    \end{itemize}
\end{itemize}

Note in generel that the main point of our paper is, that the {\bf two numbers
in the third column agree} for most of the scales. The agreement for the
susy unification is though not so impressive. This means that our idea
with the fluctuating lattice does {\bf not agree well with supersymmetric
  SU(5) unification !}
However, the minimal SU(5) unification and the inflation Hubble constant
$H_{inflation}$,
seems to fit better. The to the Hubble Lemaitre expansion rate at inflation
associated $V^{1/4}$ we only get into our scheme
by using its relation to this infaltion Hubble Lemaitre expansion in the
inflation time by LFRW relation. Therefore we wrote ``consistency'' for this
$V^{1/4}$ case. 

\begin{figure} 
  \includegraphics[scale=0.7]{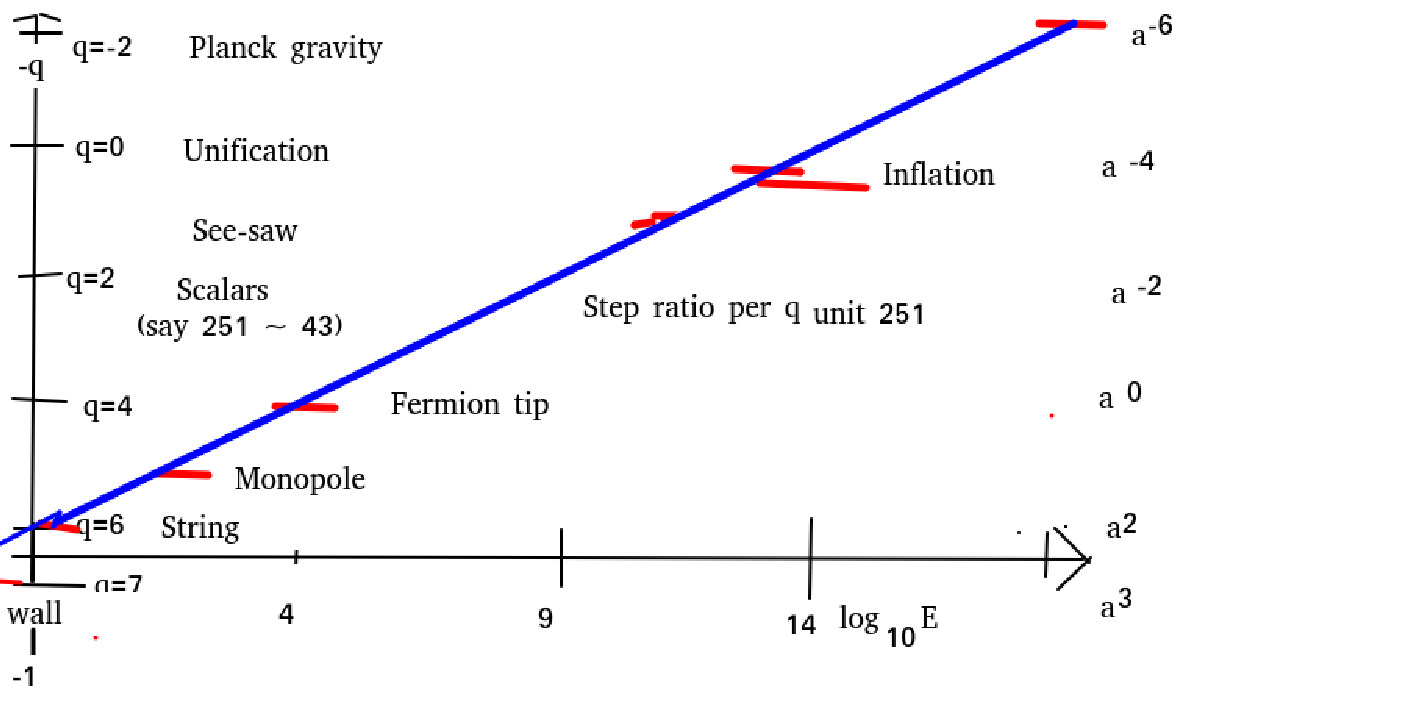}
  \caption{\label{all} Here we see how the energy scales $E$,  along
    the abscissa, the logarithms of them,  (for easiness of calculation
    we used basis 10 logartihm). They all fit within expected accuracy
    except as it is put here the ``inflation scale'' for which is it not
    so obviuos what to take for $q= 4+n = 4+\hbox{``the power for a relevant
      for averaging''}$. In the table we decided that there were two scales
    involved with inflation, the Hubble-Lemaitre expansionrate H and the
    fourth root of the potential for the inflaton. For the first $a^{-4}$
    and for the second $a^{-5}$ was chosen, and then it fits well.
    The ordinate is the number $q=n+4$ or the power $n$ for the
    to be averaged $a^n$. The assumed Galton distribution for the
    link length $a$ gives the straight line as prediction.
    For the energy scale called ``scalars'', which is my invention just as
    a scale at which there would be many scalar particles having their masses
    and therefore also some vacuum expectation values of scalar fields
    like the Higgs (but the known Higgs is {\em not}
    there
    ! (see subsubsection \ref{Higgsmass}) of the order of
    this
    ``scalars'' scale. The insertion on the plot of ``(251 $\sim$ 43)'' alludes
    to that the step in energy $E$ per step in the power $n$ is 251 while a
  typical suppression of fermion masses by an extra charge seperating the left
  and the right Weil components is estimated to 43.(see subsection
  \ref{scalars} or \cite{Universe} for
  details)
  }
    
\end{figure}

\begin{figure}
  \includegraphics[scale=0.7]{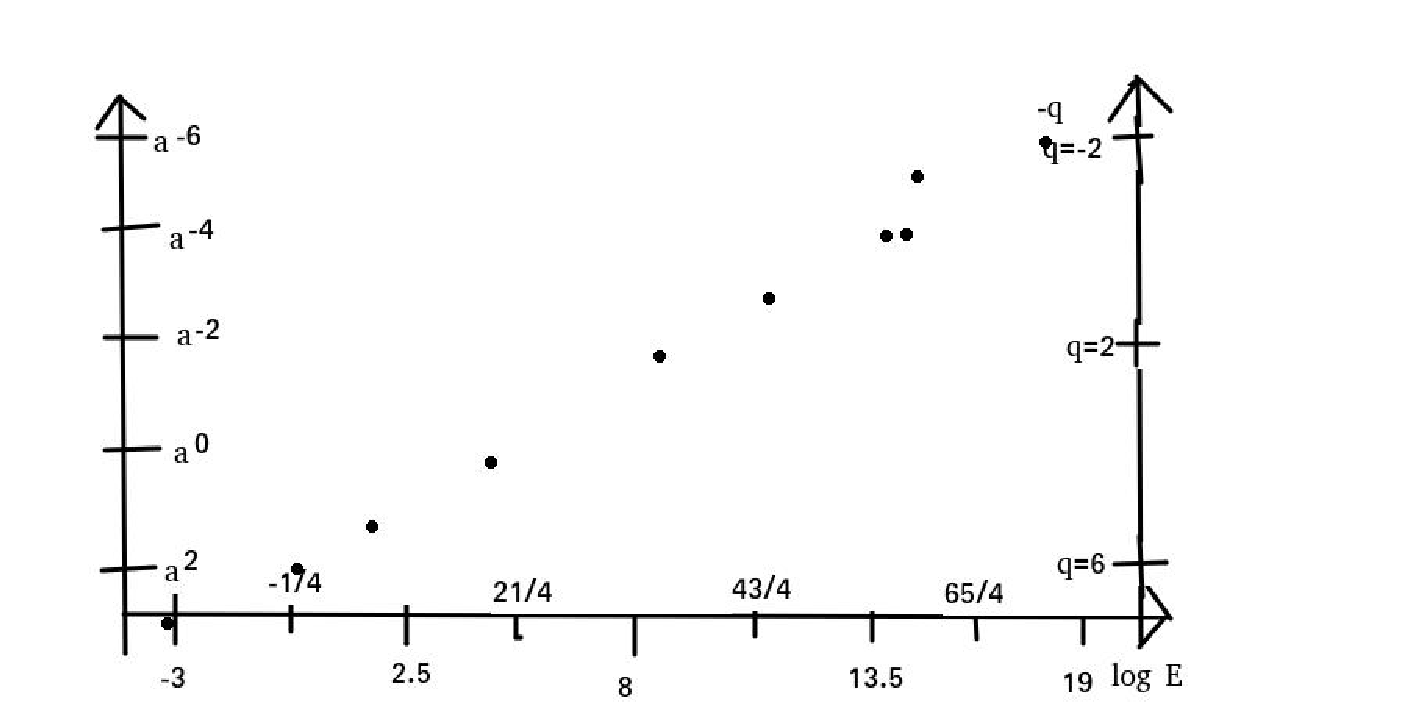}
  \caption{\label{wd} Similar figure as
    figure \ref{all} but only with reduced Planck scale
    for the Planck scale, and written with just dots to let us
    enjoy how ell it fits the straight line. The cosmological entries
    looks worst and susy grand unification was left out.}
\end{figure}
\subsubsection{Standard Model Higgs needs a finetuning somehow or another}
\label{Higgsmass}
              One needs to have finetuning making the known Higgs
      so light as the 125 Gev, and that must
      overwrite, that the scalars should have had the ``scalars scale'' masses.
      Personally I would have to think of our theory with complex action
      and influence from the
      future\cite{CAT, CAT2,Coincidensies,Relation} as giving some
              fine tuning mechanism which could overwritting the
              suggestion that the scalrs should have mass at the
              ``scalars scale'' give the much smaller Standard Model Higgs
              mass 125 GeV.

\subsection{On the Scalars and the Seesaw scales and the Little
  hierarchy problem} \footnote{The words ``little hierarchy problem'' is
    used as the left over of the hierarchy problem after helped by susy,
    but we here (miss)use the words in the meaning of the wondering: Why
    are the
    quarks and leptons
  not having masses of the same order of magnitude?}
\label{scalars}

The see-saw- scale is simply the mass order of magnitude for the
see-saw neutrinos which in the see-saw model cause the neutrino
oscillations of the observed Standard model neutrinoes. There are
probably several such see-saw-neutrinoes and in the picture we shall take
here they have order of magnitudewise the same masses, namely the
see-saw-energy-sclale as mass. To determine this mass scale requires
trusting some detailed theory for the see-saw neutrinoes and their
couplings, but looking at the most trustable theoretical physicists
proposals, we extract a scale for the see-saw masses of the order
of $10^{11}$ GeV, which fit our straight line wonderfully. By analogy
we now propose, that there should be a similar energy scale called
``scalars'', at which there are several by order of magnitude similar
scalar boson masses. Because of the mass square going into the
Lagrangian for the scalar bosons while it is the mass itself that
goes analogous in for the fermions, we find that in our scheme of
energy scales the mass scale for scalars should be one factor
251 lower than that for the fermions, the see-saw- neutrinoes.
This 251 is the step in energy scale corresponding to one unit
in $q$ or equivalently in the power $n$ in $a^n$. But now in the philosophy
of all couplings being a priori of order unity scalar field having
vacuum expectation values will have them of the order of
the ``scalar scale'' .

But then from the point of view of the see-saw fermions - which have masses
of the see-saw scale - the vacuum expectation values of the scalar bosons
possible spontaneously breaking various symmetries would look like, that the
{\bf very weak breakings} because the scalar scale is lower and the
breakings are of the scalar scale order of magnitude. So here popped out
of our scale hierarchy a mechanism for getting weak symmetry breakings without
having to put in further small numbers. All the couplings we still assume
of order unity. But such {\bf weak} symmetry breakings is exactly, what
can be usefull to make the hierarchy of fermion masses in the Standard
Model by claiming the right and left Weyl components of the various
Standard Model fermions being typically diffeerent with respect to some
of these weakly broken charge types. It is exactly such weakly boken charges
that may explain the so called small hierarchy
problem\cite{FroggattNielsen, qarketcmasses}
of explaning the
relativly large mass ratios among the Standard Model fermions.

But now in \cite{Universe} and in the present article we predict the typical
suppression of a fermion mass by having an extra charge needed to be broken
to have an effective  mass term  to be a factor equal to the ratio of the
see-saw scale over the scalar scale. In estimating this typical suppression
factor of the masses in the Standard Model fermion spectrum in the article
\cite{Universe} we got something of order 43, while the ratio of the two
energy scales see-saw versus scalars scale should be about 251 from our fitting.
But the deviation between 43 and 251 is not great, the uncertainties of such
estimates taken into account. So we consider this story of there being
scalar scale as successful.

\section{The anomalous magnetic moment deviations}
As a little progress in the last moment before submission of the manuscript to
the Corfu-proceedings let us make an attempt to see, if there is any
chanse for our scheme with fluctuating lattice to give good numbers
for the deviations of the experimental anomalous moments for muon and
electron from the theoretically calculated values. It should namely be had
in mind, that while one might usually expect, that the genuine cuttting off
new physics might wait till the Planck scale, then the situation in our
present scheme gives much better chanse for seeing indeed the new physics
providing the cut off effect, that is at the end needed for the quantum field
theories to converge, to make sense. Our scheme suggests the most fundamental
scale to be the ``fermion tip'' one,and that is formally already inside the
experimental  reach 10 TeV!.

\subsection{Plan and Simplification to come through at first.}

For simplicity we shall not truly calculate the corrections from our
fluctuating lattice, since even the lowest order diagram is
quite complicated. Instead we shall choose a typical integral, that
must occur in the evaluation of the finite lowest order anomalous magnetic
moment contribution, and look for choosing such a representative
integral so as to arrange our calculation of the correction by our
(lattice) cut off 
as easily calculable as possible.

\subsection{Choice of Representative Integral}

We basically choose the form of the integral we would get
by ignoring in the lowest order diagram - or
putting to zero - the external four momenta and keeping only the
denominators in the propagators. Since we are going to look for a
relative correction, we do not have to care for the overall
normalizaton. The divergent integrals in the true calculation of the
$g-2$, i.e. anomalous magnetic moment, should cancel out, so we coose our
representative integral to be convergent.

Indeed we choose to imagine, that we have already made the Wick rotation and
denoting the numerical magnitude 
of the loop 4-momentum by $q$. Then we propose the following
``representative integral'':
\begin{eqnarray}
  \int_0^{\infty} \frac{q^3 dq}{(q^2+m^2)^2q^2}&=&
  \frac{1}{2} \int_0^{\infty} \frac{d(q^2)}{(q^2+m^2)^2}\\
  &=&\frac{1}{2} \int_{m^2}^{\infty}\frac{dY}{Y^2}\\
  &=& \frac{1}{2}[-\frac{1}{Y}]_{m^2}^{\infty}\\
  &=& \frac{1}{2m^2}
  \end{eqnarray}

\subsection{Introducing our Lattice Cut off}

Now we shall a bit crudely introduce a cut off very similar
to that of a fluctuation lattice. But really we do not believe in the
cut off in Nature being exactly a lattice cut off; so rather we would
like to just modify the propagators in the perturbation theory in a
nice smooth way, and admit that we do not really know exactly how
this modification should be, except that we replace the $q$ or $q^2$ in the
denominators of the propagators by some smooth - analytical -
function, which is not very special, so that its Taylor expansion has
coefficients of order of magnitude unity. In order to think of a concrete
modification we replace the size of the four-momentum $q$
(thought after the weak rotation, so that this size $q$ is real and
positive) by $\frac{1}{a}sinh(a q)$, which has the same property
of being odd under sign change for $q$ as $q$ itself has, and
which has such a sign of the Taylor expansion coefficient for
$q^3$ that the propagator immediately begins to deviate from the
usual un-cut-offed propagator $\frac{1}{m^2+q^2}$ in the right direction to
make the
propagator $\frac{1}{m^2 + \frac{1}{a^2}\sinh^2(a q) }$ rather quickly
disappear, as it has to at high energies in order
to work as a cut off.
We make thus the replacement of  representative $sinh$-type:
\begin{subequations}
\begin{eqnarray}
  q &\rightarrow & \frac{1}{a}\sinh( a q)\\
  &\approx & q + q\frac{1}{6}(a q)^2 +...\\
  &\approx & q (1+\frac{1}{6}(a q)^2)\\
  \hbox{thus } q^2 &\rightarrow & q^2( 1+ \frac{1}{3} (a q)^2)\\
  \hbox{and } m^2+q^2  &\rightarrow& (m^2+q^2)
  (1 +\frac{1}{3}(a q)^2*\frac{q^2}{m^2+q^2})\\
  \frac{1}{q^2(q^2+m^2)^2}&\rightarrow&\frac{1}{q^2(q^+m^2)^2}*
  \left (1-\frac{1}{3}(a q)^2(1+2\frac{q^2}{m^2+q^2})\right )\\
  \hbox{but integration measure } q^3dq &\rightarrow& q^3dq\;\; \hbox{is NOT
    changed}\\
  \hbox{So }\int \frac{q^3dq}{q^2(m^2+q^2)^2}&\rightarrow&
  \int \frac{q^3dq}{q^2(m^2+q^2)^2}\left ( 1
  - \frac{1}{3}(aq)^2(1+2\frac{q^2}{m^2+q^2})\right )\\
  &\approx&  \int \frac{q^3dq}{q^2(m^2+q^2)^2}\left ( 1
  - (aq)^2\right )\hbox{ for large $q$ in correction}\nonumber
\end{eqnarray}
\end{subequations}

We here took the attitude that it is only the propagators
which are modified by the cut off, while the integration measure
$\int ...d^4q \propto  \int ...q^3dq$ is just kept.
Had we made a true lattice cut off, we should have made a sharp cut off in the
integration region, but with our $sinh$-type we have gotten an exponential
cut off at high $q$ values, and we can let it take over the role of the
cut off in the integration region, and thus truly keep the
unchanged integration measure. We have indeed replaced the true lattice
cut off by a more general type of cut off, which is easy by the
$q$ integration measure being kept undisturbed (i.e. no sharp cut off).

Let us illustrate the by our choice a very easy calculation first:
the large $q$ approximation for the correction due to the cut off:
\begin{subequations}
\begin{eqnarray}
  \hbox{``correction''} &=&- \int \frac{q^3dq}{q^2(q^2+m^2)^2}*(aq)^2\\
  &=&- \frac{1}{2}\int\frac{d(q^2)}{(m^2+q^2)^2}*(aq)^2\\
  &=&- \frac{a^2}{2}\int\frac{((q^2+m^2) -m^2)d(q^2+m^2)}{(m^2+q^2)^2}\\
  &=&- \frac{a^2}{2}\int_{m^2}^{\infty}\frac{(Y-m^2)dY}{Y^2}\\
  &=& -\frac{-a^2}{2}+\frac{a^2}{2}\int_{m^2}^{\infty}\frac{1}{Y}dY
\end{eqnarray}
\end{subequations}
We found here a logarithmically divergent bit, but that was because we
used the approximation that $q$ was much smaller than the genuine cut off
at $aq$ of order unity. So we shall simply cut the logarithmic divergence
off at
\begin{eqnarray}
  Y&=& q^2+m^2 = (\frac{O(1)}{a})^2+m^2\\
  &\approx& \frac{O(1)}{a^2} \hbox{ for high cut off}
\end{eqnarray}

If we which to calculate as accurately as we can in a primive way, we
we could say, that the order of magnitude quantity, $O(1)$ denoted here,
should be chosen so, that at the point $a q=O(1)$ the cut off correction factor
$\frac{\sinh(a q)}{a q}$ be of order say 2. i.e.
\begin{eqnarray}
    \frac{\sinh(``O(1)''}{`` O(1)''} &\approx & 2,
    \end{eqnarray}
so that the propagator or square root of it will be damped to
about the 1/2.

Since
\begin{eqnarray}
  \frac{\sinh(2.18)}{2.18} &=& 2.00,
  \end{eqnarray}
we should then take the suggestive order of unity quantity to 
\begin{eqnarray}
  ``O(1)''&=& 2.18. 
\end{eqnarray}

So the logarithmic term becomes
\begin{eqnarray}
  \hbox{``log-term''} &=& -\frac{a^2}{2}\int_{m^2}^{\infty}\frac{dY}{Y}\\
  &=&- \frac{a^2}{2}*\ln(\frac{
    2.18
  }{(am)^2})
\end{eqnarray}

The finite term just deviates from original integral
$\frac{1}{2}\int\frac{d(q^2)}{(m^2+q^2)^2}$ by an extra factor
$m^2a^2$, so that
\begin{eqnarray}
  \hbox{`` finite-term''} &=& \hbox{``original''}*(m^2a^2)\\
  &=& \frac{1}{2m^2}*(m^2a^2)\\
  &=& \frac{1}{2a^2}.
\end{eqnarray}
Since we have been completely careless about the normalization, we have now
to concentrate on the relative correction due to the cut off:
\begin{eqnarray}
  \hbox{``Relative correction'' }&=& \frac{\hbox{``Log-term''+ ``finite-term''}}
    {\hbox{``original''}}\\
      &=&- (am)^2\ln(\frac{2.18}{(am)^2})-(am)^2\\
      &=&- (am)^2\ln(\frac{2.18}{(am)^2*e})\label{first}.
\end{eqnarray}

It is obviously seen, that this expression in first approximation
goes as $a^2$ as function of the link size $a$, and thus
according to our table the value to be used for the
cut off energy $\sqrt{<a^2>}^{\; -1}$ about the string-scale,
which we interprete as hadrons being strings, and the hadronic scale
\begin{eqnarray}
  \hbox{the string energy scale }&=& 0.16 GeV \hbox{(fitted)}\\
  *-*-*-*-*-&*&-*-*-*-*-*\nonumber\\
  \hbox{For muon $\mu$: } \sqrt{<(a^2 m_{\mu}^2)>}&=& \frac{106 MeV}{160 MeV}\\
  &=& 0.66\\
  \ln(0.66) &=& -0.41\\
  <(a m_{\mu})^2>&=& 0.66^2= 0.44\\
  *-*-*-*-*-&*&-*-*-*-*-*\nonumber\\
  \hbox{for electron $e$; }\sqrt{<a^2 m_e^2>}&=& \frac{0.51MeV}{160Mev}\\
  &=& 0.0032\\
  \ln(0.0032) &=& -5.75\\
  <(a m_e)^2>&=& 0.0000102=1.02*10^{-5}\\
  *-*-*-*-*-&*&-*-*-*-*-*\nonumber\\
  \hbox{for tau $\tau$ ; } \sqrt{<a^2 m_{\tau}^2>}&=& \frac{1777MeV}{160 MeV}\\
  &=& 11.1\hbox{ (not much sense)}\\
  \ln(11.1)&=& 2.41\\
  <(a m_{\tau})^2> &=&11.1^2 = 123
\end{eqnarray}
But this argumentation is only true provided the the two factors $a$ in the
correction causing quantities $(a m)^2$ are indeed the same.{\bf So we shall
  not trust the just above equations}.

\subsection{Alternative to just using $<(a m)^2>$, namely $<a m>^2$}

If we use the $<(a m)^2>$ as suggested above without fluctuating
lattice, it means that we assume, that the two $a$-factors in the
expression $<(a m)^2>$ really are the local lattice scales $a$ from
{\bf the same point in the fluctuating lattice}. But is that really true?
In the just above discussion we got the appearance of the second power in
$a$ by insisting on a rotational invariant propagator, so that
especially we should end up with a propagator, which after having been made of
the
form $\frac{1}{m^2+q^2}$, should keep the denominartor being of {\bf even}
order in $q$. We therefore imposed our choice for the modification
of the propagator to be so that it replaces odd order in $q$ expressions
with odd order in $q$ and even order ones by even order. This, however,
likely is a too strong restriction of, what the cut off could do enforced
only cut-off caused modifications of second or at least even order in
in $q$ and thus in $a q$ to come in.

But that may mean imposing the fluctuating lattice too much
symmetry. Especially if we think of the $q$ as
a true four momentum, rather than just as the norm of such a four vector,
a cut off caused replacement by a lattice would  not even keep translational
invariance a priori, but only after averaging out at  the
end, it would  be
the simple  propagator $\frac{1}{q^2+m^2}$.

In fact we could think something like this:
\begin{eqnarray}
  q &\rightarrow& q(1+a q)\\
  \frac{1}{m^2+q^2} &\rightarrow&\frac{1}{m^2+q^2(1+2a q)}\\
  \hbox{With averaging} &\approx \rightarrow& \frac{1}{m^2+q^2+ 2q^2 <aq>}
  \end{eqnarray}

If you require the end result to be both translational and  rotational
invariant, all the terms
linear in $<a q>$ disappear and only by going to second order in
$<a q>$ you hat any correction from the cut off. You must calculate with
second order terms in $<a q>$ included and discard by averaging under rotations
all the fisrt order terms in $<a q>$.

Order of magnitudewise you get at the end of having imposed rotational
invariance much the same as before, but now we got the change
\begin{eqnarray}
  <(a q)^2> & replaced\;  by& <a q>^2. 
\end{eqnarray}
For a narrow distribution of $a$ this would make
no difference, but for the Galton distribution (\ref{Galton}) we get a
widely different result.

In fact with the $<a q>^2$ we have the typical energy scale to use to be the
``monopole scale'' according to our table. That is to say
\begin{eqnarray}
  <a >^{-1} &\approx & 40 GeV \hbox{(the fitted value)}  
\end{eqnarray}

(The last paragraphs I should excuse for not being convincing nor
pedagogical, but even if you just take it to mean: Now we allow ourselves
to adjust which of the two energy scales, the ``string scale'' $\sim 0.16 GeV$
or the ``monopole scale'' $40 GeV$ or $27 GeV$, to use at
the two different places in the very simple formula (\ref{niceformula}),
the overall factor and the place in the logarithm.

Even then, it would mean, that we fitted the two anomalous magnetic moment
deviation by selecting one possibility among $2\times 2=4$ possibilities. If
we take
into account, that we also has chosen, whether to believe the finestructure
constant
from the Cs or the Rb measurement, we would have chosen among 8 possibilities,
but still it is not so much!)

\subsection{The Observed Deviations}

We here give a table with calculations meant to check a couple versions
of our model(calculation) for the anomalies in the anomalous magnetic moments
$a=\frac{g-2}{2}$, for the three charged lepton, but of course the
$\tau$ magnetic moment is essentially not measured and of course not with
the needed accuracy for seeing deviations - here denoted $\Delta a$ from the
in Standard Model predicted value. We have here taken the sign of this
deveiation $\Delta a$ as {\bf the experimental value minus the theoretical one.}

\begin{adjustbox}{width=\textwidth}
\begin{tabular}{|c|c|c|c|c|}
  \hline
  1.&  text&electron&muon&tau\\
  \hline
  \hline
 2.& Experimental anomalous MM.&$a_e^{exp}$=0.00115965218059(13)&
  $a_{\mu}^{exp}$ = 0.0011659209(6)& $a_{\tau}^{exp}$= 0.0009 $\pm$ 0.0032\\
  \hline
 3.& SM theoretical anomalous MM.&$a_e^{SM}$=0.001159652181643(764) &
  $a_{\mu}^{SM}$=0.00116591804(51)&$a_{\tau}^{SM}$=0.00117721(5)\\
  \hline
 4.& Deviation, exp - SM :&$\Delta a_e(Cs) = (-101\pm 27)*10^{-14}$
  &$\Delta a_{\mu} = 249(22)(43)\times 10^{-11}$& \\
&  &$\Delta a_e(Rb)=(34\pm 16)*10^{-14}$&&\\
  \hline
  5.&Relative deviation &$\frac{\delta a_e(Cs)}{a_e} = (-87\pm 23)*10^{-11}$&
  $\frac{\Delta a_{\mu}}{a_{\mu}}= 214(19)(37)*10^{-8}$&\\
 & &$\frac{\Delta a_e(Rb)}{a_e} = (29\pm 14)*10^{-11}$&&\\
  \hline
  6.&Our $<(a m)^2>$ & $<(a m_e)^2>= 1.05*10^{-5}$& $<(a m_{\mu})^2> = 0.44$&
  $<(a m_{\tau})^2> =123$ (nonsense ?)\\
  \hline
7.&  $\frac{\hbox{Relative dev.}}{<(a m)^2>}$ &
  $\frac{\Delta a_e(Cs)}{a_e<(a m_e)^2>}=(-83\pm 22)*10^{-6}$&
  $ \frac{\Delta A_{\mu}}{a_{\mu}*<(a m_{\mu})^2>} = 486*10^{-8}$&\\
 & &$\frac{\Delta a_e(Rb)}{a_e<(a m_e)^2>}=(28\pm 15)*10^{-6}$& &\\
  \hline
  8. &Alternativ: $<a m>$& $<a m_e>$=$1.27*10^{-5}$& $<a m_{\mu}> 2.7*10^{-3}$ &
  $<a m_{\tau}> = 4.5*10^{-2}$ \\
 & $<a m>^2$ &  $<a m_e>^2 = 1.63*10^{-10}$ & $<a m_{\tau}>^2 = 7.0*10^{-6}$ &
  $<a m_{\tau}>^2 = 2.0 *10^{-3}$\\
  \hline
  9. &$\frac{Relative\;  dev.}{<a m>^2}$ &$\frac{Rel. \; dev._e(Cs)}{<a m_e>^2}=
  -5.34 \pm 1.41$&$\frac{Rel. dev._{\mu}}{<a m_{\mu}>^2}=0.306(21)(53)$ &\\
 & &$\frac{Rel. \; dev._e(RB)}{<a m_e>^2}= 1.78\pm 0.85$&&\\
  \hline
    \end{tabular}
\end{adjustbox}

\subsubsection{The different lines in the table:}
\begin{itemize}
\item[1. line] The headings for the columns.
\item[2. line] The experimental values of the anomalous magnetic moments
  of the charged leptons.
\item[3. line pair] The theoretical Standard Model anomalous magnetic
  moment values
\item[4.line pair](Here we include two lines formally intoline 3.) The
  deviation of the experimental anomalous magnetic moments from the
  Standard Model calculation one. But now an extra line not counted is
  sneaked in because the electron anomalous magnetic moment is measured
  so accurately that a 5 s.d. tension in the determination of the fine structure
  constant between measuring using Cs and Rb becomes important, so we must
  give the deviation for the twodifferent values of the fine structure
  constant.
\item[5. line pair] The deviations from 3. $\Delta a$ divided by the value
  of the anomalous magnetic moment itself for the lepton in question.
\item[6. line] Our from our in the table fitting to
  $<a^2> =\frac{1}{\hbox{``string energy scale''}^2} $ multiplied by the
  square of the mass $m$ of the lepton announced in the top line. Since this
  mass $m$ is just a constant of course
  \begin{eqnarray}
    <(a m)^2> &=& <a^2>*m^2 = \frac{m^2}{(0.16 GeV)^2}.
  \end{eqnarray}
\item[7. line pair] In this line pair we divided the relative deviations
  from line pair 4. by the averages in line 5., because if it really had
  the right thing to do have these $<(a m)^2>$ from line 5. provide the
  small size of the cut off effect, so that it was correct in (\ref{first})
  to put $(a m)^2$ equal to $<(a m)^>$, we should have gotten these ratios
  in this line pair 6. close to unity. But that seems not at all to be right.
\item[8. line] We then attempt to instead put the over all
  factor in (\ref{first}) $ (a m)^2$ equal to $<am>^2$, which is rather
  \begin{eqnarray}
    <a m>^2 &=& <a>^2*m^2= \frac{m^2}{\hbox{``monopole scale''}^2}
    =\frac{m^2}{(40GeV)^2}.
  \end{eqnarray}
\item[9. line pair] When we now divide line pair 4., the relative deviations
  by the line 7. then we get indeed numbers of order unity. So if we can argue
  that the two $a$ factors in front in (\ref{first})arenot correlated
  and shall be averaged separately, then these numbers in this line pair
  8. being of order unity is a success of our model, (the log factor
  naturally shall be of order unity)
\end{itemize}

\subsection{Backward fitting}

If we take the philosophy, that the momentum scale, at which the cut off
roughly sets in is given by the momentum scale, at which
\begin{eqnarray}
  <(a q)^2> &\approx & 1\\
  \hbox{meaning that at } q=q_{cut \; off} &\approx& \frac{1}{\sqrt{<a^2>_{formal}}}
  = \hbox{``string energy scale''}= 0.16 GeV\\
  &&\hbox{(as if the two $a$'s at same
    point)},\nonumber\\
\end{eqnarray}
we might argue for using the two $a$ factors in $<(a q)^2>$ to be at the same
point/hypercube for the cut off in the momentum going into
the logarithm by thinking of it as a higher order term due to the
lattice of the lattice Lagrangian term for the kinetic energy tern
$F_{\mu\nu}F^{\mu\nu}$ say. Both factors $a$  have to be at the
hypercube of the contribution to the $F_{\mu\nu}F^{\mu\nu}$.
This should give that the $<(a m)^2>$ in the logarithms in (\ref{niceformula})
should from the ``string scale''.

But if we imagine, that the lattice takes up even momentum for small moments,
although it delivers it back to keep phenomenologically/at the end momentum
conservation,
a term like $(a m)^2$ would after being average to $<(a m)^2>_{true}$, we call
it, really have two uncorrelated factors $a$, meaning from
different sites, and the value would be rather
$<(a m)^2>_{true} = <am>^2$ (meaning use ``monopole scale'').

\begin{eqnarray}
  \hbox{Thus truly rather } <a^2>=<a^2>_{true} &\approx& <a>^2\hbox{(when
    $a$s at different points)}\nonumber\\
   && \hbox{ because the two $a$'s are often  different}\\
    \hbox{and thus } \frac{(<a^2>_{formal})^{-1}}{(<a^2>_{true})^{-1}}
    &\approx& \left(
    \frac{\hbox{``string energy scale''}}
         {\hbox{``monopole energy scale''}}\right )^2\\
         &\approx & (\frac{0.16 GeV}{40 GeV})^2,
  \end{eqnarray}
we might absorbing our order of unity factors into
the ``string energy scale'', by putting
\begin{eqnarray}
 (``string'')^2 &=& \frac{2.18}{e}* (\hbox{``string energy scale''})^2,
  \end{eqnarray}
   get the very simple form of our prediction for the deviation of the
  anomalous magnetic moment from the Standard model value:

  \begin{eqnarray}
    \frac{\Delta a_l}{a_l} &=&- \frac{m_l^2}{\hbox{``monopole energy scale''}^2}
    *\ln(\frac{\hbox{``string''}^2}{m_l^2})\\
    &=&2\frac{m_l^2}{\hbox{``monopole energy scale''}^2 }
    \ln(\frac{m_l}{\hbox{``string''}}).\label{niceformula}
    \end{eqnarray}

  Of course having only measured two anomalous magnetic moments,
  it is not so great to fit with two parameters, ``string''  and
  `` monopole energy scale'', except if we find that the
  fitting values agree remarkably well  with these scales as
  found by our fitting in the table above.

  We may simply extract the fitting parameters by using
  \begin{eqnarray}
    \frac{\Delta a_e}{a_e*m_e^2}- \frac{\Delta a_{\mu}}{a_{\mu}m_{\mu}^2}
    &=& \frac{2 \ln(m_e/m_{\mu})}{\hbox{``monopole energy scale''}^2}.\nonumber\\
    \hbox{meaning for, say ``Cs'': } &&\nonumber\\
    \frac{(-87\pm 26\%)*10^{-11}}{(0.000511GeV)^2}-
    \frac{214(9\%)(17\%)*10^{-8}}{(0.105658GeV)^2}&=& \frac{2\ln(
      \frac{0.000511GeV}{0.105658GeV})}{\hbox{``monopole energy scale''}^2}
    \nonumber\\
    \hbox{or } (-3.332\pm 26\%)*10^{-3}GeV^{-2}
    - 1.917(9\%)(17\&)*10^{-4}GeV^{-2} &=&
    \frac{2*(-5.332)}{\hbox{``monopole energy scale''}^2},\nonumber \\
    \hbox{so } \hbox{``monopole energy scale''}
    &=& \sqrt{\frac{-10.664}{(- 3.523*10^{-3}\pm 26\% )GeV^{-2}}}\nonumber\\
    &=& 55.0 GeV\pm 13\% \label{fit55}\\
    \hbox{to compare with } && 40 GeV \hbox{ or }27 GeV.
    \end{eqnarray}

  Next we use say
  \begin{eqnarray}
    \frac{m_{\mu}}{\hbox{``monopole energy scale''}} &=&1.920 *10^3\\
    \hbox{and thus }  \ln(\frac{m_{\mu}}{\hbox{``string''}})&=&
    \frac{\Delta a_{\mu}}{a_{\mu}}/2*\frac{\hbox{``monopole energy scale''}^2}{
      m_{\mu}^2}\\
    \frac{214(19)(37)*10^{-8}}{2*(1.920*10^{-3})^2}
    &=& 0.290(9\%)(17\%)\nonumber\\
    && \hbox{(taking ``monopole energy scale'' as exact)}\\
    \hbox{this means }\frac{m_{\mu}}{``string''} &=& \exp(0.290) = 1.34\\
    \hbox{thus } \hbox{``string''}&=& 0.105658GeV /1.34\\
    &=&0.0790 GeV\\
    \hbox{to compare with }
    \hbox{``string''} &=& \frac{2.18}{e}*0.16 GeV\\
    &=& 0.128 GeV
    \end{eqnarray}
  It deviates by only a factor 1.6. The to the anomalies in the
  anomalous magnetic moments fitted ``string energy scale'' would be
  0.10 GeV.

  So this choice of making the $<(am)^2>$ in the logarithm the
  ``string energy scale'', while the occurance in the overall
  factor be ``monopole energy scale'' fits wonderfully the small
  anomalous to the anomalous magnetic moments! It should be stressed
  that these ``energy scales'' are gettable from our fit of the
  scales to the straight line, so even if only some of the energy scales,
  we dream about, are truly physical we can get this, using the straight
  line fit,
  just found wonderful
  numbers for the deviations from the anomalous magnetic moments from the
  Standard Model values. It must though be admitted that we so to speak
  predict the negative deviation for the electron anomalous magnetic moment
  and thus predict that it must be the Cs experiment value of the fine
  structure constant, that is the right one. However, for the muon anomalous
  magnetic moment the `` string energy scale'' and the muon mass are so
  close, that the sign of the logarithm of these two numbers becomes uncertain,
  and therefore we do not predict the sign of the muon anomaous magnetic moment
  deviations; but we can certainly fit it to be positive, as it is.

  \section{Conclusion}

  We have put forward a list of energy scales, all in principle
  determined from some measurements although heavily having to be
  treated by some theory to be determined, and found that they all
  fall on a straight line in a certain plot. This plot has on one axis
  the power of the lattice length $a$, i.e. $n$ in $a^n$, which is
  relevant for the energy scale in question. On the other axis we
  have simply the logarithm of the energy of the energy scale.

  Under the speculation
  of there existing a fluctuating lattice - meaning  a say Wilson lattice,
  but with the important feature, that it is somewhere very tight and
  somewhere with very big masks, and even extended somewhere in some directions
  and somewhere in other directions - we (only to order of magnitude,
  but still some fitting) have fitted these different energy scales
  by arguing, that they result from averaging the different powers of the
  link length, which are relevant for the different scales. The point is
  that with the extremeliy broad distribution of
  link sizes, which we assume - a Log Normal distribution (sometimes
  called a Galton or a Gibrath distribution) with a large width, $\sigma$ -
  the energy
  extracted from
  the average of different powers of the link length $a$ can differ by
  appreciable
  factors. In fact the energy scale extracted from the average of the
  $n$th power, $<a^n>$, is for dimensional reasons $\sqrt[n]{<a^n>}^{\; -1}$, and
  one calculate in the Log Normal distribution trivially, that such energy
  scales become of the form
  \begin{eqnarray}
    \sqrt[n]{<a^n>}^{\; -1} &=& a_0^{-1}\exp(-\frac{\sigma *n}{2})
  \end{eqnarray}
  where there are only the {\bf two} parameters, $a_0$ and $\sigma$,
  fitting order of magnitudewise the whole series of about 9 different
  energy scales, all related to some experimental data, although often
  in a very theory dependent way. These theories  are not very trustable,
  rather a kind of phenomenological models. Nevertheless it is not so bad
  again:
  
  The ``Fermion tip scale'' can in principle be considered an extrapolation
  of the fermion masses in the Standard Model put up after their number
  after mass counted from the highest mass. The extrapolation then
  back to number zero gives as the extrapolated mass what we call the
  ``fermion tip scale''; it turns out to be a parabolic extrapolation and thus
  not so accruate (for the minmum point,which is ``fermion tip scale''), but
  in principle it is defined.

  The energy scale, at
  which the running fine structure constants are closest to be in the ratios
  as if we had SU(5) unification, is also order of magnitude-wise
  making sense, and of course the Planck scale is order of magnitude-wise
  well-defined. Similarly, if we think of hadrons being approximately
  described by a string theory - as historically was the
  starting point for string theory\cite{histstring} - the order of magnitude for, say,
  the
  string tension is pretty well defined.

  So at least the energy scale values of  four of our nine scales are not so
  ill defined.

  If you believe, that the doubtfull dimuon resonance with mass
  $27 GeV$ really is a monopole for some Standard Model gauge group
  $S(U(2)\times U(3))$
  or a bound state of some such monopoles, of course this mass $27 GeV$
  points out a well defined energy scale, so if you counted that, we would
  have even 5 numerically sensible energy scales.
  
  {\bf Even having five energy scales with reasonably well-defined
    energy numbers, when plotted with the logarithm of the energy
    number versus the supposed relevant power $n$ of the lattice length,
    fall on a stright line is quite a remarkable coincidence
    unless it has some explanation.} Of course the fluctuating
  lattice is the sort of explanation which is called for.

  It is, however, quite intriguing, since while the approximate
  SU(5) unification scale and the Planck scale are easily imagined
  to come from some fundamental lattice or other type of cut off,
  the hadron string scale is basically known to be given from
  the running of the QCD strong coupling $\alpha_S$ and thus should
  not be given a priori from mysterious phantasy lattice (fluctuating
  or not). Similarly, even if the domain wall of Froggatt`s and mine
  dark matter model existed, then it should presumably be understandable
  as e.g.
  QCD vacuum  having several phases appearing, when considering the vacuum
  as function of the quark masses (`` Columbia Plot''\cite{Columbia, Columbia2}). But that would
  again be given by
  QCD and seemingly have little to do with a fundamental lattice, even if there
  were such one.

  The remaining 4 energy scales are, however, only estimable by
  using some phenomenologigal theory fitting the related data
  and comming out with some number from that fitting: E.g.
  you can theoretically fit neutrino oscillations by many
  models, and these different models can make the typical or average
  see-saw neutrino have by orders of magnitude deviating values, so that
  a see-saw-neutrino-mass scale is uncertain with a few orders of
  magnitude. Similarly the inflation time Hubble-Lemaitre constant
  is quite model dependent, and the domain wall tension in our
  dark matter model is very badly determined, even if our dark matter 
  model should be true.

  The least trustable of the scales is the ``scalars scale'',
  which is only my own speculation: I propose that there should be an
  energy scale,
  for which there are several scalars having their mass, and if they
  have expectaion values in vacuum also the scale of these expectation values.
  Nevertheless even this worst among the nine scales is formally
  involved with an experimental connection in as far as some typical
  scales for ratios of standard model Fermions can be used to fix it. In fact
  since the fermions in the Satandard Model deviate from each other by
  big factors of the order of 100 or more, explaining this the effect of
  some fermion masses needing breaking of more approximately conserved
  charges of some type yet to be discovered than other fermions in Standard
  Model, is a natural idea. Now such only weakly broken symmetries
  could appear, when fermions at the see-saw scale of mass have symmetries
  broken by scalar vacuum expectation values of size given by the
  ``scalars scale''. This scale is namely by our straight line fit
  lower than the ``see-saw scale'' by a factor say 251. 
  The ratio of the
  ``see-saw scale'' to the ``scalars scale'' should thus be a typical ratio
  of Standard Model Fermions, and thus in principle some experimental
  connection is there. In fact in\cite{Universe} I looked up for some
  more than once occuring ratio values of pairs of Standard model
  Fermions and found:
  \begin{eqnarray}
    \frac{m_t}{m_b} \approx \frac{m_b}{m_s}\approx \frac{m_s}{m_u} &\approx& 43
    \label{s43}\\
    \hbox{or if Yukawa coupling $=\sqrt{4\pi}$,} \frac{``see-saw''}{``scalars''}
    &\approx & 43*3.54 = 152\label{s152}.
  \end{eqnarray}
  
  \subsection{The anomalous magnetic moments}
  
  All this was already in the Universe article\cite{Universe}, but
  a quite new section, also not present at the time of the conference,
  was the attempt to estimate the deviation of the anomalous magnetic
  moments of electron and muon from Standard Model values, which our
  fluctuating lattice
  would give to these anomalous magnetic moments. Our estimation
  here is, however, very preliminary in the sense, that we only
  take a ``representative integral'' for the finite integral
  supposedly providing the lowest order loop integral for the anomalous
  magnetic moment. But the result for fitting the deviations of the
  anomalous magnitic moments may not depend so much on the
  exact ``representative integral'', since our calculation turned out to
  be mainly
  a logarithmically divergent term cut off in the high energy end by the
  cut off loop-momentum identified (by our assumption) with
  scale of the ``string'', which is modulo order unity the scale
  of ``hadronic strings''. We could namely then hope, that many different
  barely finite integrals going into the anomalous magnetic moment calculation
  would by being modified by some factor loop momentum squared $q$ would
  also go logartihmically divergent in a very similar way, so that
  the difference in the relative correction would be small. The over all
  scale/strength of the cut off
  effect is by our calculation and assumptions, however, the scale
  we call in the table the ``monopole scale''. That it is such a rather simple
  logarithmic divergence cut off makes it likely that the various
  types of integrals for which we should have calculated the correction
  to get the true
  correction to the full anomalous magnetic moments, would likely all be
  corrected by the same factor. If so, then we should already
  have performed the correct calculation.

  Our model predicts that the correction to the electron anomalous
  magnetic moment  should be negative in the sense of diminishing
  experimental value 
   compared to the theoretical anomalous magnetic moment.
  Thus our model has only a chanse of being right, if we use the
  value of the fine structure constant determined by use of
  caesium Cs, while the one using rubidium leads to no significant
  deviation of theoretical and experimental magnetic moments for the electron.
  However, the sign of the muon anomalous magnetic moment is due to an
  accident not comming out of our model, but the order of magnitude 
  of the deviations from theory of both the anomalous magnetic moments
  - electron and muon - come out surprisingly well. The coincidense, which
  makes the muon anomalous magnetic moment have the unpredictable
  sign , is that, what we call the hadron string scale, which is
  given as 0.16 GeV in our table, is very close to the muon mass, which
  is 0.105658 GeV. It is namely the logarithm of the ratio of these
  two numbers corrected by some order unity factor that determines the
  sign of the muon anomalous magnetic moment contribtuion from our cut off.

  So the our model has the small deviations 
  of theory from experiment for the anomalous magnetic moments as a great
  success, although we must admit that it is an a bit uncertain discussion
  why one should use for the overall magnitude of the deviations the
  ``monopole scale'' in stead of e.g. also the ``string'' scale.
  But if one does not use the monopole scale one gets completely wrong
  magnitude for the deviations.

  We think this calculation of the anomalous magnetic moment
  corrections in our model deserves more precise calculations.

  If our speculation that one of our energy scales is one
  of monopole masses - probably some monopoles for the gluon fields
  and presumably what was found at the mass 27 GeV decaying into  two muons
  were not a genuine monopole, but rather a ``hadron like'' bound state
  of a couple of monopoles, so that no net monopole charge would prevents
  its decay into two muons. But one could still ask, if such monopoles
  could have an important role for confinement? One could ask:

  Could it be that confinement is not appearing for truly continuous
  gluon fields, but that it is only  possible to have confinement,
  if there effectively are some monopoles, like on a lattice
  as used in lattice calculations, or some physically existing
  gluon-monopoles, even if they should be as heavy as of the
  order of 27 GeV?

\section{Appendix on the Significance for Gravity}

\subsection{Abstract of the appended note on Gravity}
{This contribtution is only meant as a little appendix to the
  contribution ``Approximate Minimal SU(5), Several Fundamental Scales,
  Fluctuating
  Lattice''\cite{ThisCorfu} telling about, what this work means for
  especially gravity.
  Indeed the most common lattice size in the fluctuating lattice, which
  is our model here, is hugely larger than the Planck length, and so only
  in a very tiny fraction of space time the links are so small as the
  Plack scale. But the important regions, where the energy momentum tensor is
  felt by the gravitational fields, are only, where the fluctuating lattice
  happens to have exceedingly small link sizes, much closer to the Planck
  length. When as in our model here the world is governed by the
  (fluctuating) lattice on much bigger length scales than the gravity
  Planck length, we tone down the importance of gravity for the most
  fundamental physics.
}




\subsection{Introduction to the little Gravity Note}

In both my talks I mainly talked about the proceedings contribution
``Approximate Minimal SU(5), Several Fundamental Scales, Fluctuating
Lattice''\cite{ThisCorfu}(a review of\cite{Universe},which developped
from \cite{AppSU5}), in which I point out that a series of energy scales
associated
with different branches of physics, such as the Planck energy scale
associated with gravity, an {\bf approximate unified scale} at which the running
gauge coupling constants in the notation for SU(5) unification are most close,
and say a scale I call ``Fermion tip scale'', which is an extrapolation of the
Standard Model fermion masses to most heavy such fermion (number 0 from the
top), ...can be represented on a straight line. This is in a plot in which we
along one axis has the power into which the link length $a$ should be
put before being averaged to extract  the energy scale in question, and along
the other axis the logarithm of the energy of the energy scale. When I here
talked about taking an average, e.g. $<a^n>$, of a power of the link size $a$
it was meant with a fluctuating lattice in mind, so that in some regions
the link size is very small, while in others it is very large.

\subsection{On Gravity in Our Scale Model}

In the present note it is our intension to put out some thinking about,
what our model with the fluctuating lattice is supposed to mean for gravity
interaction:

The surprise of the fitting to the straight line mentioned is, that
if we count the average size of all the links in the truly existing
lattice in physical space time, it is supposed that the inverse of this
typical link size is of the order of the ``fermion tip energy scale'',
which is only about $10^4$ GeV. This $10^4$ GeV is a hugely lower energy than
the Planck scale, supposed to be the energy scale representing gravity.
Actually new lattice physics at only $10^4$ GeV should be much
more hopeful to be seen experimentally than Planck scale physics, which is
usually supposed to be connected with quantum gravity. But in our model
it means, that quantum effects for gravity first will be important
for so high energy, that the lattice has completely taken over the
usual quantum field theory. We shall indeed imagine in our picture, that
the interaction between the gravitational fields, vierbeins or $g^{\mu\nu}$,
and matter essentially only takes place along regions, where the
density of hypercubes is exceedingly high and the length of the links
correspondingly minute. But regions with so small links will only be tiny
stribes in the space time. Alone the picture of the in this sense
gravitational interaction taking place essentially only tiny regions
gives an intutiive feeling for the weakness of the gravitational force.

That there is at all in our model calculable gravitational interactions
(the Plack scale) is due to that we have assumed a definite link size
distribution - a log normal one - so that the amount of the very tiny
region essentially dominating the provision of gravitational interaction
is assumed via this distrubtuion.

If we assume, as is what one usually hopes for fora lattice, that
lattice cuts off the loop divergensies in the field theories, such divergencies
will be cut off at much lower enrgy scales than the Plackscale already, and
specalive physics at Plack scale or higher will be completely modified
by and berried in the lattice physics.

Thus the gravity will be appreaciably toned down in the search for the
fundamental physics.

We would indeed have lattice effect non-locallities at much longer length scales
than the Planck scale.

Indeed at the end of the long contribution\cite{ThisCorfu} to this Corfu
Institute
we have a section on the anomalies in the anomalous magnetic moments for
electron and muon. Accepting our stories about the modification of the
anomalous magnetic moments due to our fluctuating lattice and the
parameters taken from the fitting to our series of the different energy scales
with our straight line we can get good values for the deviation from the
standard model calculations without having to put in further parameters!

You can take this succes in predicting modifications to the main loop
integral for the anomalous magnetic moments as having already ``seen''
the nonlocality effects of the lattice.

\subsection{Conclusion for Gravity Appendix}

We have in this note argued, that taking serious our fluctuating
lattice model one might tone down the problems of quantizing gravity,
in as far as one should at much lower energy scales than the Plack scale
see effects of the lattice. 

We can claim that our, still too preliminary though, estimate of the small
deviation of the anomalous magnetic moments for elctron and muon is a
first looking at the cut off scale, being much closer than most poeple
previously thought.


\section*{Acknowledgement}
  The author thanks the Niels Bohr Institute for status as emeritus.
This work was discussed in both Bled Workshop and the Corfu Insitute
this year 2024.

\end{document}